\documentclass[pre,twocolumn,floatfix,superscriptaddress,10pt]{revtex4-1}
\usepackage{graphicx,amsfonts,amssymb,amsmath, hyperref,multirow,array}
\usepackage[utf8]{inputenc}

\AtBeginDocument{\renewcommand{\natexlab}[1]{#1}}

\newif\ifhyper
\hypertrue
\ifhyper
\hypersetup{
  citecolor = {green},
  colorlinks = {true}, 
  urlcolor = {blue} 
}
\fi

\newlength{\ldag}
\begin{document}

\title{Granular rheology: a tale of three time scales}

\author{O. Coquand} 
\email{olivier.coquand@univ-perp.fr}
\affiliation{Institut f\"ur Materialphysik im Weltraum, Deutsches Zentrum f\"ur Luft- und Raumfahrt (DLR), 51170 K\"oln, Germany}
\affiliation{Laboratoire de Mod\'elisation Pluridisciplinaire et Simulations, Universit\'e de Perpignan Via Domitia,
52 avenue Paul Alduy, F-88860 Perpignan, France}

\author{W. T. Kranz} 
\email{kranz@thp.uni-koeln.de}
\affiliation{Institut f\"ur Theoretische Physik, Universit\"at zu K\"oln, 50937 K\"oln, Germany}
\affiliation{Institut f\"ur Materialphysik im Weltraum, Deutsches Zentrum f\"ur Luft- und Raumfahrt (DLR), 51170 K\"oln, Germany}

\author{M. Sperl} 
\email{matthias.sperl@dlr.de}
\affiliation{Institut f\"ur Materialphysik im Weltraum, Deutsches Zentrum f\"ur Luft- und Raumfahrt (DLR), 51170 K\"oln, Germany}
\affiliation{Institut f\"ur Theoretische Physik, Universit\"at zu K\"oln, 50937 K\"oln, Germany}



\begin{abstract}
	We adapt statistical models of the physics of complex fluids to study the rheology of granular liquids.
	This allows us to provide laws of granular rheology based on first principles, which compare well with
	previously established phenomenological laws.
	In particular, the very successful law of $\mu(\mathcal{I})$ rheology can be understood within our model
	as the lowest order non trivial Pad\'e approximant of the macroscopic laws of rheology if one takes into account
	processes taking place at three distinct types of time scales: Collisions occurring at microscopic scales, collective motions like cage effect taking
	place at intermediate, mesoscopic scales, and finally advection that takes place at the macroscopic time scale.
	Our model's ability to describe granular physics outside of the Bagnold scaling regime
	allows for a natural extension to the rheology of granular suspensions.
\end{abstract}

\maketitle

\section{Introduction}

	Granular fluids are omnipresent in our everyday life.
	The study of their behavior is important for many industrial applications, but it is also crucial to the understanding of
	some geological processes such as avalanches \cite{Savage79,Savage89,Savage98,Pouliquen99,Pouliquen02}, pyroclastic and debris
	flows \cite{Kelfoun08,Kelfoun09,Gueugneau17,Ogburn17,Salmanidou17}, and sediment transport \cite{Frey09,Frey10,Pahtz20}, as well
	as gravisensors in plants \cite{Berut18,Forterre18}, and specific animal behavior \cite{Ruhs20}.
		
	Despite the absence of attractive force in the simplest granular flows, three distinct flow regimes can be identified
	depending on the granular fluid's density \cite{Andreotti13}: at low density, collisions are relatively scarce, this is
	the \textit{gaseous} regime; at higher densities --- typical packing fractions $\varphi$ in the range
	$0.4 \lesssim \varphi \lesssim 0.6$ --- the grains experience
	very frequent collisions, which significantly affect their qualitative behavior, this regime is called the \textit{liquid}
	regime; finally, close to the jamming transition, interparticle friction becomes relevant with deep consequences
	\cite{Staron10,DeGiuli16,DeGiuli17a,DeGiuli17b}.
	Importantly, in the intermediate liquid regime, some microscopic characteristics of the granular particles
	become irrelevant \cite{Tapia19}, which allows for the existence of universal laws.
	Most examples of granular flows on Earth are in the liquid regime \cite{Andreotti13}, this study focuses on this latter one.

	When a granular liquid is in the Bagnold flow regime --- which is generally the case when no external source of driving
	power other than shear is present --- its rheology is described by a phenomenological law, called the $\mu(\mathcal{I})$ law,
	that has been determined by fitting a huge data set including flows in numerous geometries \cite{GDR04}.
	This law describes the evolution of the effective friction coefficient $\mu$ of the granular liquid,
	a quantity that qualifies how far from a simple liquid the system lies --- in simple liquids the effective friction is weak ---
	as a function of a dimensionless version of the shear rate, called the inertial number $\mathcal{I}$, which compares the
	shear rate to the typical time scale of the motion of particles between collisions.
	The granular liquid regime roughly corresponds  to the range $0.05\gtrsim\mathcal{I}\gtrsim 0.003$.
	The limiting value correspond respectively to the onset of the friction dominated regime for the lower bound,
	and the breakdown of the continuum medium approximation to describe the granular fluid for the upper bound.
	Since then, the $\mu(\mathcal{I})$ law has been tested against even more data, from a wide variety of flow configurations
	(from a simple shear experiment to the collapse of a granular column), and has shown a remarkable agreement with the
	experimental and numerical data both at the qualitative and the quantitative level
	\cite{DaCruz05,Jop05,Cassar05,Jop06,Pouliquen06,Forterre08,Peyneau08,Lagree11,Tankeo13,Clavaud17,Fullard17,Fullard19,Delannay17,
	Pahtz19}, even in the most recent studies \cite{Tapia19}.
	This formula still has one weakness however; it remains so far only phenomenological \cite{Forterre18}, the physical
	origin of this simple rheology has not been found out yet.

	In a recent study \cite{Coquand20}, it has been shown that models inspired from the theory of complex liquids can
	be successfully adapted to describe granular flows.
	More precisely, it showed that the so-called Granular Integration Through Transients (GITT) formalism
	--- that describes the rheology of frictionless dissipative hard spheres ---
	provides a set of fundamental equations which, once numerically solved, yield results showing a satisfactory
	agreement with the predictions of the $\mu(\mathcal{I})$ law, with parameter values compatible with the experimental
	results.
	An explicit confirmation of this has been given in a recent series of experiments \cite{Angelo23}.
	However, the set of GITT equations is very complex, and can be solved only numerically.
	The purpose of this paper is to build simpler toy-models in which the rheology of granular
	liquids is explained as a competition between three time scales associated with the relevant physical processes at play in the
	system --- collisions, shear advection and structural relaxation --- occurring at the respective microscopic, macroscopic and mesoscopic scales.
	This model not only allows to retrieve the $\mu(\mathcal{I})$ law within a theoretical framework with a well identified set
	of hypotheses as a lowest order Pad\'e approximant of our macroscopic rheology,
	but also gives non-trivial predictions as for the behavior of the effective friction coefficient $\mu$
	--- a central rheological quantity --- outside of the Bagnold regime where most experiments are conducted,
	thereby allowing to understand a wider range of rheological behaviors.
	As a result, the model can be easily generalized to the case of high density granular suspensions, which provides a model
	for the evolution of $\mu$ in a regime where the search for such a law is under active investigation
	\cite{Courrech03,Cassar05,Boyer11,DeGiuli15,Guazzelli18,Tapia19,Pahtz19,Suzuki19}.

	The paper is organized as follows: in the first section, we present the toy-model and show that general properties
	of granular liquid flows can be explained through the competition between two time scales.
	Then, in a second section, we introduce the third time scale, derive the evolution of the effective friction 
	coefficient, and generalize the model to granular suspensions, identifying the various flow regimes through the
	relative strength of the involved time scales.
	Finally, we conclude.

\section{The Two time scales Toy model}

	Before entering the details of the toy-model, let us first recall how liquid state theory can be adapted to capture the granular liquid
	phenomenology.

	\subsection{The Granular Integration Through Transients formalism}

		Let us consider a granular liquid consisting of $N$ infinitely hard particles,	
		of restitution coefficient $\varepsilon$ and granular temperature $T$.
		For the sake of simplicity, we restrict ourselves to the case of an incompressible planar shear flow.

		The main challenge in describing the dynamics of granular liquids is the fact that, because of the dissipative character of the
		collisions, the system behaves generally not as a Newtonian, but as a complex fluid, as we are going to show below.
		This means that, depending on the conditions in which the liquid evolves, the relation between the shear stress and the strain rate may
		not be linear.
		Such an effect has to be captured already at the level of the equations of motion that govern the dynamics of the fluid.
		Moreover, since the type of stress-strain rate relation depends a lot on internal, structural characteristics of the liquid,
		the equation of the dynamics is established at the level of functions of the internal structure of the liquid, rather than at the level
		of the particle of fluid (in which case the information about its internal content would be lacking).
		
		The dynamics of the system is taken to be given by the Mode-Coupling Theory (MCT), which accounts for the slow down
		of the relaxation of correlation functions due to the cage effect caused by clogging of particles at high density
		\cite{Goetze08}.
		The general form of the MCT equation is that of a Mori-Zwanzig equation for the dynamical structure factor $\Phi_q$,
		which is nothing but the normalized density correlation function in Fourier space:
		$\Phi_q(t)=\left<\rho_q(t)\rho_{-q}\right>/S_q$,
		$S_q=\left<\rho_q\rho_{-q}\right>$ being the static structure factor.
		The general form of this equation is given below:
		\begin{equation}
		\label{eqMCT}
			\begin{split}
				\ddot{\Phi}_{q}(t) &+ \nu_{q}\dot{\Phi}_{q}(t)+ \Omega_{q}^2 \Phi_{q}(t) 
				+ \Omega_{q}^2 \int_0^t \!\!d\tau\,m_q(t-\tau)\dot{\Phi}_{q}(\tau) =  0 \:.
			\end{split}
		\end{equation}
		The detailed expressions of the various coefficients appearing in this equation can be found in the 
		appendix \ref{AMCT}. We chose not to reproduce them here since their lengthy expressions do not really impact our reasoning.
		The first three terms of Eq.~(\ref{eqMCT}) describe a simple relaxation of $\Phi_q(t)$ controlled by the two
		characteristic frequencies $\nu_q$ and $\Omega_q$, as for simple liquids.
		They express the weakening of the initial correlations through time and space~: As time grows, two particles that were close to each
		other at an initial time will on average be separated by an arbitrarily large distance, namely, after a sufficient amount of time,
		particles in a liquid loose the information about their neighbors.
		These are the terms dominating at moderate enough densities where the granular medium is in the Newtonian liquid regime.
		They describe the usual interpolation between a collision-dominated, ballistic regime, at short time scales, and a large time
		diffusive regime.

		The last term accounts for the memory effects that arise when the dynamics of the liquid 
		drastically slows down at high densities, and can be expressed in the frame of the Mode-Coupling approximation.
		Physically speaking, because of the slow down of the dynamics, particles remain, on average, close together on a much longer
		amount of time, hence the time-dependent term in the equation of motion.
		In the limit where this last term dominates, the local information about the structure of the neighbors is never completely
		blurred, so that the density correlation function never decays completely.
		The expression of the quantities present in this term are not needed in our derivation, and presented in appendix \ref{AMCT}.
		For details, the reader is referred to the previous papers on GITT \cite{Kranz18,Kranz20,Coquand20}.
	
		Whereas it has been shown that MCT tends to overestimate the importance of cage-effect in the vicinity of the glass
		transition, we are only concerned here with the dense liquid regime.
		In particular GITT assumes that shear-heating is always sufficient to make the granular material yield, so that we
		never consider a true glass phase.
		The regime of parameters under consideration is thus the one where the MCT has proven to provide an accurate description
		of the physics at play.
			
		Given the complexity of Eq.~(\ref{eqMCT}), the expression of $\Phi_q(t)$ is not known in general, even in very simple
		cases.
		In order to understand the MCT picture of the glass transition, it is useful to simplify this function thanks to
		the Vineyard approximation \cite{Vineyard58} combined with a Gaussian ansatz for the self-interacting part
		of the dynamical structure factor:
		\begin{equation}
		\label{eqVin}
			\Phi_q(t) \simeq S_q \,e^{-q^2 \Delta r^2(t)}\,,
		\end{equation}
		where $\Delta r^2(t)=\left<\mathbf{r}(t)\cdot\mathbf{r}(0)\right>$ is the mean-squared displacement (MSD).
		Hence, in the liquid phase, most particles obey a diffusive behavior of diffusion constant $D$, so that
		$\Delta r^2(t)=6Dt$, and $\Phi_q(t)$ shows a simple exponential relaxation (see Fig.~\ref{figSigmaRecap}).
		When going deeper into the supercooled regime on the other hand, the MSD develops a plateau: most particles are trapped
		by their neighbors and cannot escape a small region, this is the cage effect.
		It then follows from Eq.~(\ref{eqVin}) that $\Phi_q(t)$ also develops a plateau (see Fig.~\ref{figSigmaRecap}).
		Furthermore, as long as the system is not in the MCT glass phase, the plateau is followed by a final decay at
		later times.
		Note that in this picture the overall $S_q$ factor does not play any major role.
		
		Finally, $\Phi_q(t)$ is related to rheological quantities through the Integration Through Transients (ITT) formalism
		\cite{Fuchs02,Fuchs09}, that can be used to express the shear stress $\sigma$ and the pressure $P$ in the
		out-of-equilibrium steady
		state as integrals over the values of $\Phi_q(t)$ at former times (see \cite{Kranz20,Coquand20} for more details):		
		\begin{equation}
		\label{eqRheo}
			\begin{split}
				& \sigma = \frac{1}{60\pi^2}\int_0^{+\infty}dt\frac{1}{\sqrt{1+\frac{(\dot{\gamma}t)^2}{3}}}
				\int_0^{+\infty}dq\,F_{1}(q,t)\\
				& P(\dot\gamma) = P(\dot{\gamma} = 0) \\
				&\quad + \frac{1}{36\pi^2}\int_0^{+\infty}dt\frac{(\dot{\gamma}t)}{\sqrt{1+\frac{(\dot{\gamma}t)^2}{3}}}
				\int_0^{+\infty}dq\,F_{1}(q,t)\\
				& \quad + \frac{1}{12\pi^2}\int_0^{+\infty}dt\frac{(\dot{\gamma}t)}{\sqrt{1+\frac{(\dot{\gamma}t)^2}
				{3}}} \int_0^{+\infty}dq\,F_{2}(q,t)\,,
			\end{split}
		\end{equation}
		where $\dot\gamma>0$ is the shear rate, and the kernels in the time integrals are given below:
		\begin{equation}
		\label{eqF}
			\begin{split}
				& F_{1}(q,t) = -q^4\,\dot{\gamma}T\left(\frac{1+\varepsilon}{2}\right)\Phi^2_{q(-t)}
				\frac{S'_{q(-t)}S_q'}{S_q^2}\\
				& F_{2}(q,t) = -q^3\,\dot{\gamma}T\left(\frac{1+\varepsilon}{2}\right)\Phi^2_{q(-t)}
				\frac{S'_{q(-t)}}{S_q^2}(S_q^2 - S_q)\,.
			\end{split}
		\end{equation}
		In all these expressions, the dynamical structure factor is evaluated in a time dependent wave vector $q(-t)$.
		This is a consequence of advection caused by the shear flow: the shear flow imposes some average motion to the
		particles (with a linear velocity profile in this particular case, see Fig. \ref{figmuRecap}), which is antagonistic
		to the cage effect.
		The time integrals therefore reproduce the competition between the slow MCT relaxation and the shear advection.
		
		The limit $\varepsilon\rightarrow1$ can be taken in the above formulas to recover usual expressions in non-dissipative
		systems.
		Therefore, although we will be mostly concerned with granular flows in the following, this formalism encompasses
		the rheology of colloidal suspensions as a particular case.
		It must be noted, however, that one of the most important differences between granular flows, and those of colloidal
		suspensions is the presence of dissipative collisions in the former.
		As a result, the flowing out-of-equilibrium steady state is defined by a balance between the power injected in the system,
		and the power dissipated by the collisions, which can be summarized by the following balance equation:
		\begin{equation}
		\label{eqPowBal}
			\sigma\dot\gamma + P_D = n \Gamma_d\omega_cT \,,
		\end{equation}
		where $n$ is the liquid's density, $\omega_c$ is the collision frequency (the collision frequency can be estimated for
		example from the packing fraction by using the Enskog expression $\omega_c = 24\varphi\chi d^{-1}\sqrt{T/\pi}$ \cite{Hansen06} where
		$\chi$ is the contact value of the pair correlation function and $d$ is the diameter of the particles.
		Numerical estimates in this work, use the P[4/5] \cite{Clisby06} ansatz to estimate $\chi$,
		see \cite{Coquand20} for more details), and $\Gamma_d=(1-\varepsilon^2)/3$ is a dimensionless dissipation rate
		(see \cite{Kranz20} for more details).
		Its expression is not important here.
		Finally, $P_D$ is a generic driving term, that encompasses all sources of power injection other than shear heating.
		Eq.~(\ref{eqPowBal}) defines the granular temperature $T$.

		Let us emphasize here that $T$ is a \textit{kinetic} temperature, defined from the second cumulant of the velocity fluctuation
		probability distribution, and not a \textit{thermodynamic} one.
		In particular, due to their size, granular particles do not thermalize with their environment, even when they are in suspension.
		This point is particularly important when comparing formulas outside of the Bagnold regime in the elastic limit, where
		contrary to their granular counterpart, colloidal suspension involve the thermodynamic temperature (even if the formulas are
		the same, their physical meaning is different).
		
		All in all, in GITT the rheology of granular liquids is described in terms of integrals over the advected dynamical
		structure factor $\Phi_{q(t)}(t)$, whose dynamical evolution is described by MCT, combined with a power
		balance equation (\ref{eqPowBal}) that defines the steady state.
		This method has proven successful to describe the rheology of dry granular liquids \cite{Coquand20}.
		However, the involved structure of the equations makes it difficult to understand precisely how the underlying physical
		processes at play impact the end result.

	\subsection{Reduction of the ITT integrals}
	
		One of the main sources of complexity in the GITT equations is the coupling between the time and wave number
		dependencies of	$\Phi_q(t)$; this can be simplified.
		Indeed, in the MCT, the glass transition is described as a bifurcation process characterized by a number of
		universal quantities describing the dynamics in the vicinity of the plateau of $\Phi_q(t)$.
		It is thus possible to build a class of models, called \textit{schematic models} in which equivalent bifurcations
		apply to a function $\Phi(t)$ that is only a function of time.
		Consequently, the MCT equation Eq.~(\ref{eqMCT}) can be highly simplified, what allows for analytical studies of some
		of the asymptotic properties of $\Phi(t)$ when $t$ is very large.
		Such approach has for example been successfully applied to the rheology of colloidal suspensions
		\cite{Fuchs02,Fuchs03,Henrich05,Varnik08,Brader09},
		where it was shown that the relaxation from the plateau is dominated by the shear advection term.
		
		However, even in the simplest schematic MCT models the full time evolution of $\Phi$ does not have a simple analytical
		form.
		Consequently, we decided in this work to go even one step further and replace $\Phi_q(t)$ by a simple relaxation
		function $\exp(-\Gamma t)$, where $1/\Gamma$ is the time scale associated with the structural relaxations,
		namely the scale controlling the decay of $\Phi$ to 0.
		In the liquid phase, $\Gamma$ is typically related to the time scales appearing in the first three terms of
		Eq.~(\ref{eqMCT}), whereas when going closer to the MCT glass transition, the memory terms are more and more
		important and $\Gamma\rightarrow0$.
		This is clearly depicted on Fig.~\ref{figSigmaRecap}~: In the low density, Newtonian regime, the decay of $\Phi$
		follows a simple exponential decay with a typical rate given by the collision frequency $\omega_c$; At larger densities however,
		the decay occurs on larger time scales, namely $\Gamma$ and $\omega_c$ decouple.
		As we will show throughout this paper, the drastic simplification of our toy model is sufficient to capture the leading behavior of
		the system.
		
		By taking away the $q$-dependence of $\Phi$, we also simplify all the wave vector dependences in the integrals in
		Eq.~(\ref{eqRheo}), which reduce to mere constants.
		However, the term appearing in the integrand is not $\Phi_q(t)$ but $\Phi_{q(-t)}(t)$, and although we can safely
		ignore the wave vector dependence of $\Phi_q(t)$, the effect of advection is crucial insofar as it accounts for
		the effect of shear which is required to liquefy the system at high densities.
		Let us apply the Vineyard formula to $\Phi_{q(t)}(t)$ (in the following expression, we have used 
		the expression of the advected wave vector's norm $q(t)^2=q^2(1- (\dot\gamma t)^2/3)$ valid for the simple shear flow):
		\begin{equation}
		\label{eqVint}
			\begin{split}
				\Phi_{q(t)}(t) & \simeq S_{q(t)} e^{-q(t)^2 \Delta r^2(t)} \\
				& = S_{q(t)} e^{-q^2 \Delta r^2(t)}e^{-q^2(\dot\gamma t)^2\Delta r^2(t)/3}\,.
			\end{split}
		\end{equation}
		The time dependence appears on two levels: (i) in the static structure factor, but this effect is very mild
		compared to the drastic evolution driven by the mean-squared displacement, and
		can be safely neglected	at our level of approximation \cite{Henrich07}; and (ii) at the level of the Gaussian factor.
		The formula Eq.~(\ref{eqVint}) is useful to understand the effect of shear advection on $\Phi_q(t)$:
		close to the MCT glass transition, $\Phi_q(t)$ develops a plateau that extends over many decades in time.
		However, it is not a mere function of time, it also has a spatial structure which typically decays like a Gaussian over
		a length given by the MSD.
		When the granular medium is sheared, the advection introduces an additional time-dependence in the spatial structure
		of $q(t)$, and therefore of $\Phi_{q(t)}(t)$.
		Thus, even if the MSD were constant in the fictitious, quiescent state, the large time behavior
		of the real system would always be $\Phi_q(t)\rightarrow0$, namely, it would be shear molten.
		It should be understood that, even though Eq.~(\ref{eqVint}) is strictly speaking only valid for low
		enough values of $q$, the rapid exponential decay of the integrand always ensure that the large $q$ sector never significantly
		contributes to the integral.
		
		In our toy-model, $\Phi$ has no $q$-dependence anymore, but shear melting is required.
		Therefore, we choose to replace the advected $\Phi_{q(t)}(t)$ by the product of $\Phi(t)$ and a Gaussian screening factor
		$\exp(-(\dot\gamma t)^2/\gamma_c^2)$, where $\gamma_c$ is a typical strain scale of the system.
		Note that this is a bit different from the choice made by Fuchs and Cates in their study of colloidal suspensions
		\cite{Fuchs03},	where the advection was accounted for in the schematic model by a factor with a Lorentzian rather than
		Gaussian prefactor.
		As we are going to show in the following, the main role of the advection factor is to provide a cutoff to the time
		integral at a typical scale $1/\dot\gamma$.
		At our level of approximation, the precise form of this cutoff function is not important.
		We chose to keep the Gaussian profile because it yields simpler expressions for the ITT integrals, but as we will see, it is neither
		more nor less precise than the Lorentzian one.

		In a nutshell, the two-time-scales toy-model captures the physics of the relaxation of the density correlation function $\Phi$.
		As we argued above, this decay can occur via two competing channels corresponding to two, different, well identified physical processes~:
		The first channel is the structural relaxation with a rate $\Gamma$, corresponding to collective motion in the fluid, the other one is
		the advection channel, of rate $\dot\gamma$, corresponding to the macroscopic motion forced by the environment.

		A word of caution is in order here regarding the Gaussian screening factor.
		With this expression, the departure of $\Phi(t)$ from the plateau is quadratic in $\dot\gamma$, and not linear as most studies
		in the Mode-Coupling approximation show.
		It must be understood that, whereas most of these studies focus on the evolution of $\Phi(t)$ very close to the plateau, we 
		need here, to the contrary, to give an approximation of $\Phi(t)$ valid on the full scale of its evolution.
		In this context, a Gaussian profile that is not the most precise very close to the departure from the plateau but gives 
		a shape consistent with the full evolution of $\Phi(t)$ is a good approximation (namely, it is less precise than the schematic models
		close to the plateau, but more precise at the level of the global shape of $\Phi$).
		As can be seen on Fig.~\ref{figSigmaRecap} on the red panel, the decay from the plateau, when driven by advection, is neither
		exponential, nor Gaussian, but much faster when examined on a global scale.
		Lastly, as we already argued, the leading order behavior of the rheology is given by the location of the decay, and not the
		precise shape of the decaying function, since its role is always that of an integrand.
		
		Finally, we can simplify the fundamental ITT integrals appearing in Eq.~(\ref{eqRheo}),
		noted $\mathcal{K}_0$ and $\mathcal{K}_1$ in the following.
		The fact that we reduced the $q$-dependence leads to drastic simplifications (the square root term, even though seemingly 
		dependent on time only originates from the wave vector structure 
		\cite{Kranz20}. It has therefore been neglected as well; It is to be noted that its profile decreases anyway much slower than the Gaussian
		factor coming from advection):
		\begin{equation}
		\label{eqK0}
			\begin{split}
				\mathcal{K}_0 &= \dot{\gamma}\int_0^{+\infty}dt\int_0^{+\infty}dq\frac{F_{1}(q,t)}{
				\sqrt{1+(\dot{\gamma} t)^2/3}} \\
					      & = \frac{\gamma_c\sqrt{\pi}}{2\sqrt{2}}\mathcal{F}\left(\frac{\Gamma \gamma_c}{\dot\gamma\sqrt{2}}\right) \\
					      &\simeq \frac{\overline{\gamma}_c}{2}\,\frac{1}{1 + \overline{\gamma}_c/\text{Wi}}\,,
			\end{split}
		\end{equation}
		where $\overline{\gamma}_c=\sqrt{\pi/2}\gamma_c$, Wi$=\dot{\gamma}/\Gamma$ is the Weissenberg number, and $\mathcal{F}(x)=$ erfc$(x)e^{x^2}$.
		In this computation, the second line corresponds to the exact evaluation of the integral (let us remind that the integrand includes approximations).
		The result of this evaluation is not very useful for a comparison with experiments.
		Besides, a number of details of the variation of this function are not needed to give a faithful representation of the data.
		Therefore, we reduced, in the third line, the expression of this function to its lowest order non-trivial Padé approximant, giving a much more
		easy to use rational fraction.
		As we are going to see, this rational fraction form is largely sufficient for the need of our present study.
		It should however be kept in mind that this last step is by no mean an obvious one, since the study of sheared granular liquids in more complex
		flow geometries \cite{Coquand21}, or the study of dynamical yield surfaces \cite{Coquand23}, which are also outputs of the model,
		require a higher degree of precision.
		What we would like to put forward here is that the toy-model, however simple, gives a systematic way of building approximations that can then be
		tailored to the needs of the situation under study.

		Similarly,
		\begin{equation}
		\label{eqK1}
			\begin{split}
				\mathcal{K}_1 &= \dot{\gamma}\int_0^{+\infty}dt (\dot{\gamma} t)\int_0^{+\infty}dq
				\frac{F_{1}(q,t)}{\sqrt{1+(\dot{\gamma} t)^2/3}}\\
					      &\simeq \frac{\overline{\gamma}_c^2}{4}\,\frac{1}{1+ \overline{\gamma}_c/\text{Wi}}\,.
			\end{split}
		\end{equation}
		At this level of approximation, there is no major difference between the two types of ITT integrals.
		Details about the derivation of these formulas can be found in appendix \ref{AITT}.
		
		All in all, the complexity of GITT equations can be reduced to a simple function of one parameter, Wi,
		capturing the competition between the structural relaxation time scale $1/\Gamma$ and the shear advection time scale $1/\dot{\gamma}$
		for the control of the relaxation of the two-point density correlation function.
		This constitutes the two time scales toy model.

	\subsection{Rheology as a competition between two time scales}

\begin{figure*}
			\begin{center}
				\includegraphics[scale=0.55]{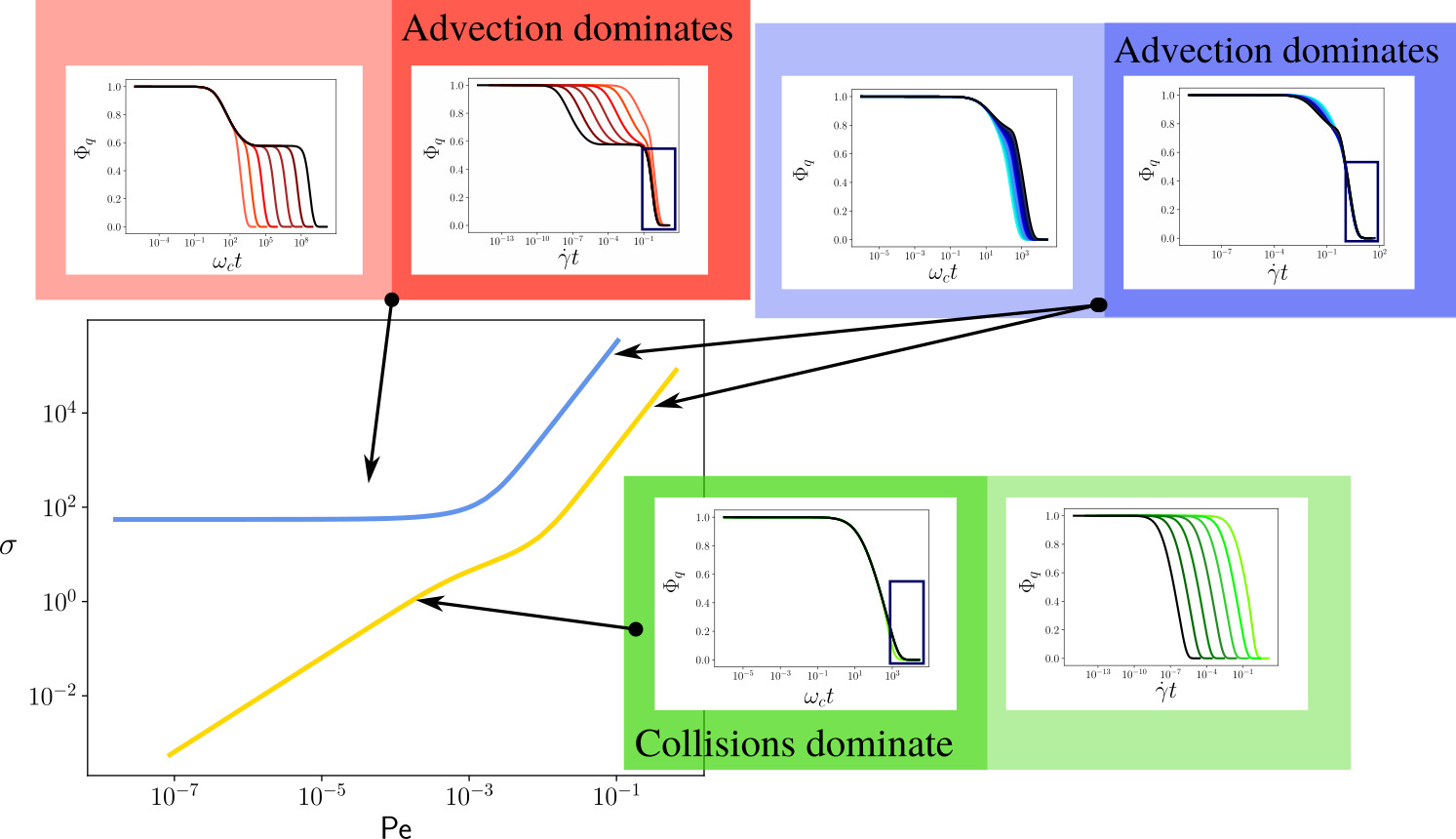}
			\end{center}
			\caption{Evolution of the shear stress $\sigma$ as a function of the Peclet number both below the MCT granular
			glass transition $\varphi\leqslant\varphi_g$ (yellow), and above it (blue).
			Three inserts display the evolution of the dynamical structure factor $\Phi_q(t)$, computed by numerically solving the GITT
			equations (\ref{eqRheo}),
			in the three different flow regimes (Newtonian $\sigma\propto\dot\gamma$ (green), yielding
			$\sigma\propto\dot\gamma^0$ (red), or Bagnold $\sigma\propto \dot\gamma^2$ (blue)).
			Each time, the left panel has a time axis rescaled by the collision frequency $\omega_c$ whereas it is rescaled
			by the shear rate $\dot\gamma$ on the right panel.
			Different curves in one insert correspond to different values of Pe smaller values of Pe corresponding to darker
			colors and larger ones to brighter colors.
			A blue rectangle indicates on which panel the different curves collapse for the final decay of $\Phi_q$.
			If the collapse is in the left panel (Newtonian regime), $\omega_c$ controls the decay; if it is in the
			right panel (yielding or Bagnold regimes), advection controls the decay.
			The green pannel corresponds to $\varphi=0.45$ and Pe$\in[10^{-9};10^{-3}]$, the red pannel
			corresponds to $\varphi=0.55$ for the same values of Pe, and the curves in the Bagnold regime have $0.45\leqslant\varphi
			\leqslant 0.58$.
			For all curves $\varepsilon = 0.85$.}
		\label{figSigmaRecap}
		\end{figure*}

		Let us examine the evolution of the shear stress $\sigma$ in the different flowing regimes, to check a posteriori the 
		quality of the approximations presented above by comparing our results to the existing literature.
		From Eq.~(\ref{eqRheo}), it is basically proportional to $\mathcal{K}_0$.
		In our toy-model, it can therefore be expressed as:
		\begin{equation}
		\label{eqSig}
		\sigma = \frac{\sigma_y}{1+\overline{\gamma}_c/\text{Wi}}\,,
		\end{equation}
		where $\sigma_y$ is some constant that accounts both for the prefactor in Eq.~(\ref{eqK0}), and a compensation for the
		$q$-dependent terms in the ITT integral.
		The competition between the times scales in Wi generates different flow regimes:
		\begin{itemize}
			\item[(i)] $\Gamma\gg\dot{\gamma}$: structural relaxation dominates. \\
				In this regime, Wi$\ll1$, therefore:
				\begin{equation}
				\label{eqNew}
				\sigma \simeq \frac{\sigma_y\,\text{Wi}}{\overline{\gamma}_c} = \eta\, \dot{\gamma}\,,
				\end{equation}
				which describes the flow of a Newtonian fluid of viscosity $\eta=\sigma_y/
				(\overline{\gamma}_c \Gamma)$.
				
				In the language of our toy-model, in this regime the shear time scale is much larger than the scale of
				structural relaxation.
				Consequently, the relaxation time depends only on the characteristic quantities of the liquid,
				and does not depend on the shear rate $\dot\gamma$.
				As an aside, it means that it does not depend either on the Peclet number Pe$=\dot\gamma /\omega_c$,
				that compares the, macroscopic, advection time scale, to the microscopic, ballistic motion time scale.
				This behavior can be checked to show up in the numerical solution of the full GITT equations displayed
				in Fig.~\ref{figSigmaRecap} in the green insert:
				on the left panel, the time axis is made dimensionless through the collision frequency $\omega_c\propto\Gamma$
				(as discussed below, this proportionality only holds in the Newtonian regime, because at low densities, the scale
				of ballistic motion, or collisions, fixes the rate of decay of the density correlation function, a relation that
				breaks down at higher densities where strongly collective motion are at play),
				all curves collapse, whereas it can be checked on the right panel, where the dimensionless time is
				$\dot\gamma t$ that different Peclet numbers are represented.

				A full account of the properties of the steady state needs to combine both Eq.~(\ref{eqSig}) and
				Eq.~(\ref{eqPowBal}).
				Injecting Eq.~(\ref{eqSig}) into Eq.~(\ref{eqPowBal}) leads to an expression $T=f(\dot\gamma,P_D)$, where $f$ is some
				function.
				Since $P_D$ is left unspecified, we can use it to tune the value of $T(\dot\gamma,P_D)$, so that in practise,
				$T$ and $\dot\gamma$ can be considered as independent variables.
				Physically, this amount to adjusting the driving of the granular material to reach a desired steady state at
				a given shear rate and packing fraction.
				This is also the origin of the third time scale compared to the well-known $\mu(\mathcal{I})$ law~: In the Bagnold regime,
				the rheology depends on a unique dimensionless number, $\mathcal{I}$, but it is also the regime where $P_D=0$; Thus,
				adding a non trivial driving power (or additional source of dissipation) introduces a new energy scale into the problem.
				
			\item[(ii)] $\dot{\gamma}\gg\Gamma$: advection dominated regime.\\
				Here, we must discriminate two different scenarios, because there are two separate causes that can
				lead the system into such a regime.
				\begin{itemize}
					\item $\Gamma\rightarrow0$: yielding regime.
					
						If $\varphi>\varphi_g$, where $\varphi_g$ is the location of the MCT granular 
						glass transition in the equivalent unsheared system \cite{Kranz10,Kranz13},
						the structural relaxations become infinitely slow (note that since the system is always shear
						molten, the existence of a true MCT glass transition, or an avoided transition with no diverging time scale,
						is irrelevant; As a matter of fact, as long as $\Gamma\ll\dot\gamma$, $\dot\gamma$ fixes the scale of the decay
						of $\Phi$ towards 0, independently of the existence of another process that may cause a decay at later time in the
						unsheared system, otherwise, the advection channel is the only relaxation channel).
						Hence, whatever the value of $\dot{\gamma}$, the condition Wi$\gg1$ is always respected.
						In that case,
						\begin{equation}
						\label{eqYield}
							\sigma\simeq \sigma_y \,,
						\end{equation}
						which is the behavior of a yielding material of yield stress $\sigma_y$.
						Let us point out that in virtue of the discussion above, even
						though Eq.~(\ref{eqRheo}) yields $\sigma_y\propto T$, using the additional degree
						of freedom $P_D$, $T$ and $\dot\gamma$ can be considered independent.
						Thus, Eq.~(\ref{eqYield}) genuinely corresponds to a shear stress independent of the
						shear rate.
						
						The corresponding evolution is displayed in Fig.~\ref{figSigmaRecap} in the
						red insert.
						Comparing the left and the right panels shows that the final relaxation time (the
						one corresponding to the decay of $\Phi$ to 0) is
						entirely determined by the Peclet number (namely by $\dot\gamma$), and does not
						depend on the collision frequency.
						
					\item Strong shear rate regime: Bagnold scaling\\
						
					Even far away from the MCT granular glass transition, it is always possible to reach
					the regime in which $\dot\gamma\gg\Gamma$ if the system is sheared strongly enough.
					The strongest shear regime corresponds to the case where $P_D=0$ in the
					power balance Eq.~(\ref{eqPowBal}), when all the injected energy is due to the shear,
					and the only source of dissipation is the dissipative collisions.
					In that particular case, the power balance Eq.~(\ref{eqPowBal}) takes a form
					called the Bagnold scaling equation\cite{Bagnold54}:
					\begin{equation}
					\label{eqBag}
						\sigma\dot{\gamma} = n \Gamma_d\,\omega_cT\,.
					\end{equation}
					Crucially, this means that $P_D$ cannot be used anymore as an adjustable parameter,
					and $T$ becomes a function of $\dot\gamma$.
					As a result, although we are still in an advection dominated regime, Eq.~(\ref{eqYield}) must be amended
					to account for the fact that $\sigma_y\propto T$ is now a function of $\dot\gamma$.
					In order to investigate this, it is thus interesting to rewrite it as $\sigma_y = \hat\sigma_y\,T$,
					where $\hat\sigma_y$ is the part of the shear stress that depends neither on $T$, nor on $\dot\gamma$.

					Since $\omega_c\propto \sqrt{T}$, Eq.~(\ref{eqBag}) yields $T\propto (\sigma\dot\gamma )^{2/3}$.
					The equation (\ref{eqSig}), thus yields:
					\begin{equation}
						\sigma = B \dot{\gamma}^2\,,
					\end{equation}
					where $B=\hat{\sigma}_y^3/\Gamma_d^2$ is the Bagnold coefficient of the granular fluid.

					Due to power balance Eq.~(\ref{eqBag}), $T\propto\dot\gamma^2$, and since $\omega_c\propto\sqrt{T}$,
					the Peclet number Pe$=\dot\gamma /\omega_c$,
					playing the role of a dimensionless shear rate, becomes constant and saturates.

					Note that while $\sigma_y$ has a very weak dependence on $\varepsilon$, the dissipation
					rate typically behaves as $\Gamma_d\propto(1-\varepsilon^2)$ \cite{Haff83}, so that $B$ is singular
					in the elastic limit.
					This should not come as a surprise.
					As a matter of fact, if the elastic limit is smooth for the Newtonian and the yielding regimes, the Bagnold
					regime requires the particular balance Eq.~(\ref{eqBag}), which can only hold if collisions dissipate energy,
					something impossible in the elastic case.
					Said otherwise, while the yielding and Newtonian regimes of rheology can be compared to their equivalent in
					colloidal systems --- up to the definition of the temperature, as argued above --- the Bagnold regime is a
					specificity of granular systems.
					
					The evolution of $\Phi_q$ in GITT in the Bagnold regime is represented in 
					the blue insert in Fig.~\ref{figSigmaRecap}.
					As expected in an advection dominated regime, the final relaxation time is controlled
					by $\dot{\gamma}$.
				
			\end{itemize}
				
		\end{itemize}

		Finally, following the reasoning of Fuchs and Cates \cite{Fuchs03} in the case of colloidal suspensions,
		we can understand the toy-model's result Eq.~(\ref{eqSig}) in the context of the viscoelastic Maxwell model.
		There is one subtlety related to the fact that the toy-model involves not only one, but two time scales:
		$\tau=1/\Gamma$ related to the structural relaxations, and $\tau_\gamma=\gamma_c/\dot\gamma$
		related to advection.
		We can use them to build a total time scale $\tau_M$ through $1/\tau_M = 1/\tau + 1/\tau_\gamma$,
		so that Eq.~(\ref{eqSig}) can be interpreted as the shear stress of a Maxwell material of shear modulus
		$G(t)=G_\infty e^{-t/\tau_M}$, with a initial shear modulus $G_\infty$ related to the yield stress
		through the following law:
		\begin{equation}
		\label{eqSy}
			\sigma_y = G_\infty \overline{\gamma}_c \,.
		\end{equation}
		Despite a different choice of advection term in the ITT integrals (Gaussian instead of Lorentzian), it is interesting
		to note that the non-linear Maxwell model of \cite{Fuchs03}, which proved successful in the description of the
		rheology of colloidal suspensions, is recovered as a particular case of our toy-model in the appropriate
		elastic limit $\varepsilon\rightarrow1$ (details in appendix \ref{ANLMM}).
		All in all, our toy model is able to describe all the known scaling regimes of granular liquid flows.

		A summary of the two-time scales toy-model can be found in Fig.~\ref{figSigmaRecap}.
		The dominating time scale ($1/\Gamma$ or $1/\dot\gamma$) determines whether collisions (Wi$u\ll1$) or advection
		(Wi$\gg1$) control the final decay of $\Phi(t)$ to 0.
		Inserts show which time scale controls the decay in three flow regimes.
		

\section{Effective friction}

	\subsection{Presentation}
	
		Granular liquids are complex liquids that share some behaviors with liquids, and other with solids.
		A useful way to quantify how far away from these two limits the system lies, is to define its effective friction coefficient $\mu$.
		This coefficient, inspired from soil mechanics, describes the ability of the system to yield in a 
		Mohr-Coulomb fashion \cite{Savage89}.
		By analogy with the Coulomb criterion of solid friction, $\mu$ is the ratio of the tangential constraint applied
		to the liquid over its normal constraint.
		In our case, it is simply $\mu=\sigma/P$.
		A small value of $\mu$ means that the system yields very easily, much like a liquid, whereas as $\mu$ gets closer to 1,
		the behavior becomes more and more solid-like.
		
		In order to determine $\mu$ in our toy-model, we need to determine the pressure.
		Following Eq.~(\ref{eqRheo}), we can decompose it as a sum of two types of terms: the unsheared pressure $P(\dot{\gamma}=0)$ which does not
		depend on advection and is therefore a mere constant (denoted $P_0$) in our toy-model, and the ITT correction given by
		the two next terms (see \cite{Coquand20} for more details).
		As discussed before, since the $q$-structure has been reduced to mere constant prefactors, both terms have the form of
		$\mathcal{K}_1$ given by Eq.~(\ref{eqK1}).
		The pressure can thus be written in a form very similar to $\sigma$:
		\begin{equation}
		\label{eqP}
			P = P_0 + \frac{P_1}{1+\overline{\gamma}_c/\text{Wi}}\,.
		\end{equation}
		In particular, deep in the liquid phase in the regime dominated by $\Gamma$, the ITT correction to the pressure is
		very weak, whereas it is stronger in the yielding regime, a feature consistent with the GITT numerical data
		\cite{Coquand20}.
		
		Finally, the effective friction coefficient can be written as follows:
		\begin{equation}
		\label{eqmu}
			\mu = \frac{M_1}{1 + M_2/\text{Wi}}\,,
		\end{equation}
		where $M_1=\sigma_y/(P_0+P_1)$ is the limit of $\mu$ in the yielding regime and $M_2=\overline{\gamma}_c\cdot
		P_0/(P_0+P_1)$.
		Hence, in the $\Gamma$-dominated regime, $\mu\rightarrow0$, whereas in the yielding regime, $\mu$ reaches a constant
		non-zero value independent of $\dot{\gamma}$ (the order of the limits is crucial here; Indeed, since the
		yielding regime is advection dominated, the inequality $\dot\gamma\gg\Gamma$ holds in any case; As a result, even when 
		$\dot\gamma\rightarrow 0 $, Wi$\gg1$ so that $\mu\simeq M_1$).
		This is all the more interesting as it has been shown that pyroclastic flows have a much lower $\mu$ than typical
		values predicted by the $\mu(\mathcal{I})$ law.
		Indeed, some processes have been suggested to explain that such flow are not in the Bagnold regime where the
		$\mu(\mathcal{I})$ law applies \cite{Gueugneau17}.
		Our toy-model confirms that some parameter ranges (corresponding to the Newtonian flow regime) are compatible
		with arbitrarily low values of $\mu$.

\begin{figure*}
			\begin{center}
				\includegraphics[scale=0.25]{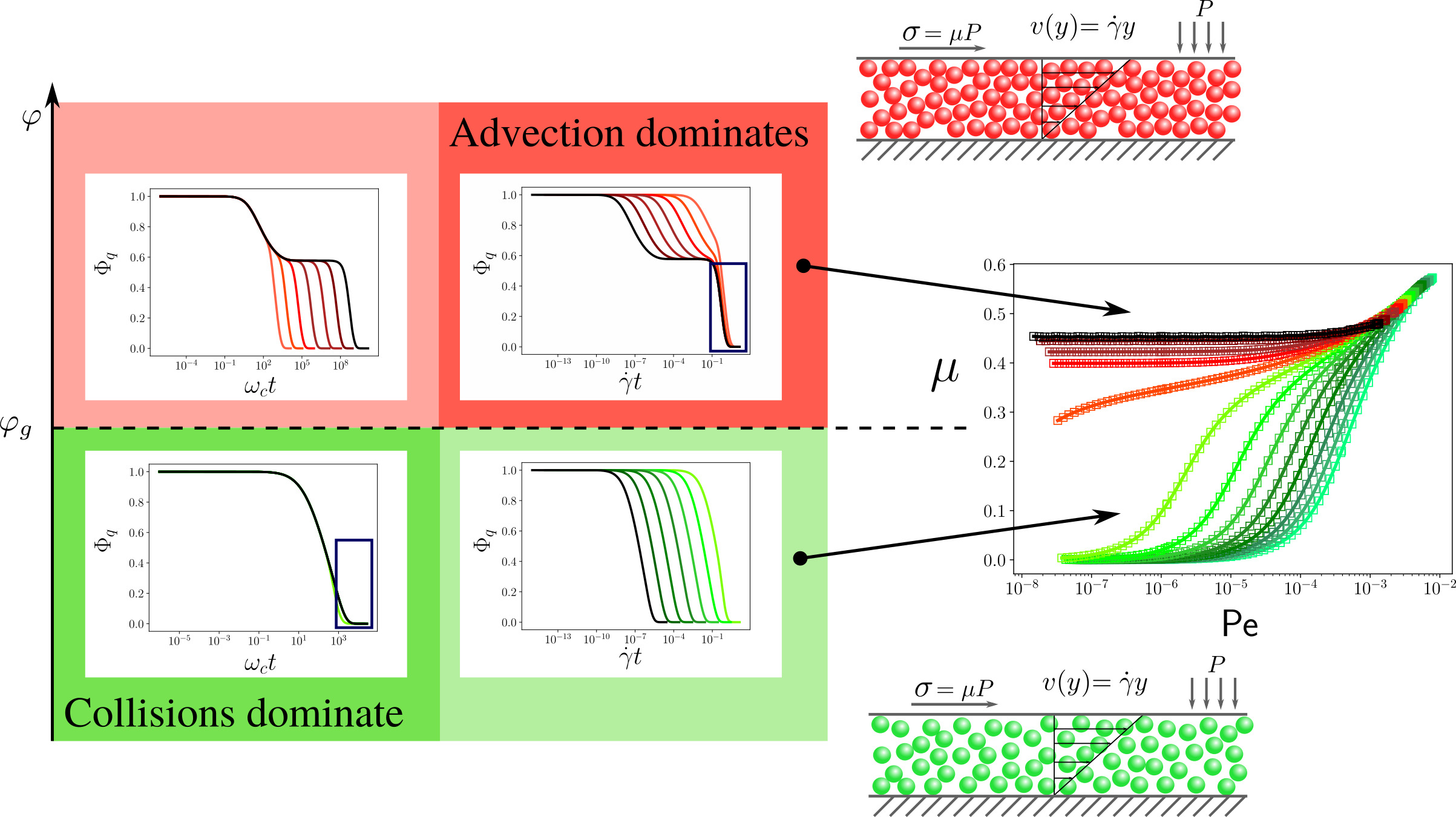}
			\end{center}
			\caption{The left part of the figures recalls the shape of the dynamical structure factor's evolution
			with time.
			On the left panel, the time axis is rescaled by $\omega_c$, whereas on the right panel, it is rescaled by
			$\dot\gamma$.
			This allows to identify the process that controls the decay of $\Phi_q$: collisions for the most dilute flows under the MCT granular
			glass transition (at $\varphi=\varphi_g$), and advection for denser flows above the transition.
			On the right side, the flow geometry is recalled, and the evolution of $\mu$ with Pe is displayed for
			various values of $\varphi$ between $0.45$ and $0.58$, for a restitution coefficient of $\varepsilon = 0.85$.
			Depending on the process controlling the decay of $\Phi_q$, the Pe $\rightarrow0$ limit of $\mu$ is either
			0 or a finite value.
			The Bagnold scaling can be observed at the level of the endpoints of the curves, for the highest
			values of Pe (see \cite{Kranz20} for more details).
			The color code on the inserts is the same as the one used on Fig. \ref{figSigmaRecap}.}
			\label{figmuRecap}
		\end{figure*}

		The predictions of the toy-model can be tested against the evolution of $\mu$ with the Peclet number Pe
		$=\dot{\gamma}/\omega_c$ computed with GITT (see Fig.~\ref{figmuRecap}).
		The following behavior is observed in the numerical data: for $\varphi\leqslant\varphi_g$, $\mu$ asymptotically goes
		to 0 when Pe decreases,	whereas it saturates to a finite value around $0.4$ for $\varphi>\varphi_g$.
		This is consistent with the prediction of the toy-model: below the MCT granular glass transition, $\Gamma$ is finite,
		and when decreasing Pe, it is always possible to reach the regime $\Gamma\gg\dot{\gamma}$ where $\mu$ can be
		arbitrarily small;
		above $\varphi_g$ however, the structural relaxations become infinitely slow, and the system stays in the yielding
		regime where Wi$\gg1$.
		At this order of approximation, our two-time-scales toy-model therefore reproduces exactly the behavior observed in
		GITT.
		
		This result is a bit disturbing though since it means that in the yielding regime, $\mu$ does not depend on Wi.
		While this seems satisfactory to describe the qualitative global tendency of the evolution of $\mu$ with Pe,
		as can be seen in Fig.~\ref{figmuRecap}, when looking at individual
		curves like in Fig.~\ref{figmufit}, $\mu$ clearly depends on Pe even in the yielding regime, even if its variations
		are much milder	(they are all the weaker that $\varphi$ is large).
		
		The origin of this is easy to understand: in the GITT curves, the behavior of $\mu$ is studied when Pe $\rightarrow0$.
		In the Newtonian regime, $\Gamma\propto\omega_c$ (no collective motion effect), so that Pe $\propto$ Wi and the identification between the toy-model
		and the GITT data is easy to make.
		In the yielding regime however, Wi$\gg1$ however small Pe is.
		This is because in this regime, the plateau in the time evolution of $\Phi$ is very long (see the upper-left quadrant
		on Fig.~\ref{figmuRecap}), the internal dynamics is very slow due to a strong cage effect, and the condition
		$\Gamma\ll\dot{\gamma}$ can be maintained even at very low values of $\dot{\gamma}$.
		The identification between Wi and Pe therefore breaks down in this regime.
		Indeed, as pinpointed in \cite{Coquand20}, the rheology of granular liquids is not defined in terms of one, but
		two dimensionless ratios of time scales: the Peclet number Pe, and the Weissenberg number Wi $=\dot\gamma/\Gamma$.

\begin{figure}
			\begin{center}
				\includegraphics[scale=0.5]{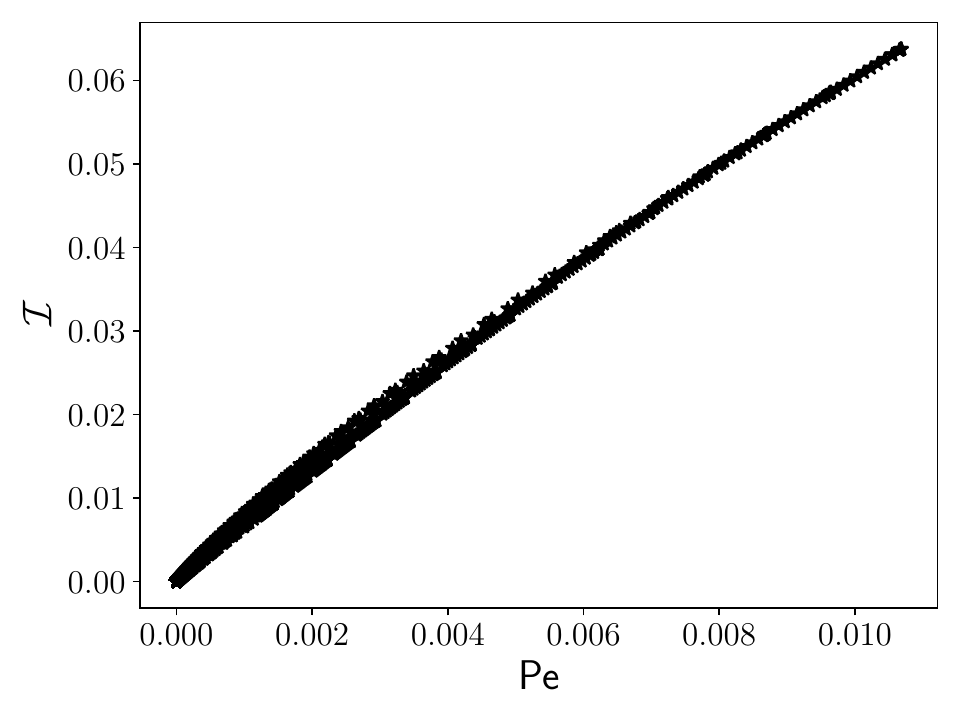}
			\end{center}
			\caption{Evolution of the inertial number as a function of the Peclet number on our dataset (all the values of Pe
			and $\mathcal{I}$ presented in all regimes in all the other figures in the paper).}
			\label{figIPe}
		\end{figure}

		Before detailing the three-time-scales toy-model, we must give a word of caution~: In our derivation, it appeared that the
		effective friction coefficient $\mu$, expressed in terms of a competition between two time scales within the two-time-scales toy-model
		does not describe well the $\mu(\mathcal{I})$ law.
		There seems to be a paradox here since by definition, $\mathcal{I}$ is a ratio of two time scales, so in the Bagnold regime, $\mu$
		does depend on the competition of two time scales only.
		However, a more careful analysis of this result shows that the Weissenberg number Wi, used in the two-time-scales toy-model is
		not a good analog to $\mathcal{I}$, it is the Peclet number that is.
		Indeed, as one can see on Fig.~\ref{figIPe}, the inertial number $\mathcal{I}$ and the Peclet number Pe are basically proportional to one
		another.
		This comes from the fact that the free-fall time scale $t_{ff}$ used in the definition of $\mathcal{I}$ is a typical time scale of the
		ballistic motion of particles, so that $t_{ff}\omega_c\simeq1$.

		Said otherwise, a proper account of the $\mu(\mathcal{I})$ rheology requires a toy-model for the general rheology with three different
		time scales (expressed in terms of Pe and Wi), such that in the Bagnold limit, one of the time scales ($t_\Gamma$ in Wi, and not $\omega_c$ in Pe)
		decouples and the Bagnold rheology can be expressed as a competition between two of the three time scales only.
	

	\subsection{The three-time-scales rheology}

		As stated before, the rheology of granular liquids depends on two dimensionless numbers: Pe that describes the
		competition between ballistic motion and advection, and Wi that in that case describe the competition between structural relaxation (including the cage
		effect) and advection.
		A consequence of the existence of three fundamental time scales --- for collisions, advection, and structural
		relaxations --- can be seen on the time evolutions of $\Phi_q(t)$ (see Fig.~\ref{figSigmaRecap} and
		Fig.~\ref{figmuRecap}).
		In the Newtonian regime, the cage effect is weak, most of the physics is captured by the competition
		between collisions and advection, and $\Phi_q(t)$ follows a simple decay.
		Closer to the MCT granular glass transition, the time scales associated with collisions and structural relaxations
		separate and $\Phi_q(t)$ follows a two-step decay.
		
		Changing from a one-step to a two-step decay can be done simply by assuming that $\Phi(t)$ does not follow a simple
		exponential relaxation, but is rather a combination of two such processes: $\Phi(t) = \lambda^{(1)} \exp(-\Gamma^{(1)}t)
		+\lambda^{(2)} \exp(-\Gamma^{(2)}t)$, with $\Gamma^{(1)}$ associated with the short-time ballistic process,
		whereas $\Gamma^{(2)}$ is associated with the long-time decay process (structural relaxations in that case, at least when they occur
		on a scale decoupled from that of ballistic motion).
		The choice of an exponential form for the first step of the decay ensures the consistency of the model in cases like the Newtonian model where
		$\Gamma^{(1)}=\Gamma^{(2)}$ and the decay occurs in one step (see Fig.~\ref{figSigmaRecap}).
		By linearity of the ITT integrals (because the term involved in the integral is $\Phi^2$ and not $\Phi$,
		the operation is not rigorously linear, but as in our model, we either face the case $\Gamma^{(1)}=\Gamma^{(2)}$ or 
		$\Gamma^{(1)} \ll\Gamma^{(2)}$, the mixed term does not play any meaningful role in the determination of the shape of $\Phi$),
		it can be checked that the resulting shear stress can be decomposed as
		$\sigma = \sigma^{(1)}+\sigma^{(2)}$, each $\sigma^{(i)}$ having the form Eq.~(\ref{eqSig}), with two respective
		time scales ratios $\Gamma^{(i)}/\dot{\gamma}$.
		Following the above discussion, the first ratio is proportional to $1/$Pe and the second one to $1/$Wi.
		The same procedure can be applied to $P$ and to $\mu$.

		The success of the two-time scales toy-model hints that the long-time relaxation process is associated with the
		more drastic variations of the rheological quantities (such as the shift from a $\mu\rightarrow0$ limit
		to a finite value of $\mu$ when Pe $\ll1$).
		In the cases where the decay of $\Phi$ is done in two well separated steps, the changes associated with the first
		step of the decay are milder, subleading variations.
		
		We therefore split $\mu$, in this fashion, introducing two separate contributions $\mu^{(1)}$ and $\mu^{(2)}$
		coming respectively from the short and the long time scales controlling the decays,
		\begin{equation}
		\label{eqmu2}
			\mu = \mu^{(1)} + \mu^{(2)} = \frac{M_1^{(1)}}{1 + M_2^{(1)}/\text{Pe}} +
			\frac{M_1^{(2)}}{1 + M_2^{(2)}/\text{Wi}}\,.
		\end{equation}
		As can be seen in Fig.~\ref{figmuRecap}, the short-time decay is completely fixed by $\omega_c$.
		In the yielding regime, the long-time decay becomes infinitely large, so that $\Gamma=0$, and Wi$\gg1$.
		Our previous paradox is therefore solved: $\mu$ does possess a contribution from yielding that stays constant
		and fixes the leading behavior, but it also encompasses a second term, due to the short-time decay, which is still
		of form Eq.~(\ref{eqmu}), and explains the remaining subleading variation.
		
\begin{figure}
			\begin{center}
				\includegraphics[scale=0.55]{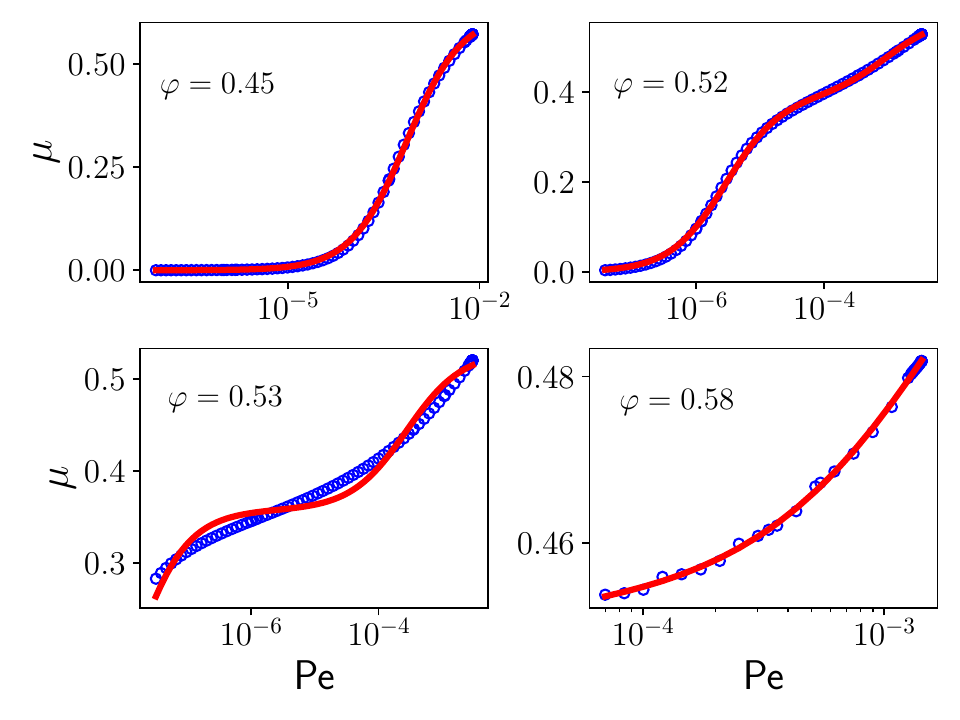}
			\end{center}
			\caption{Evolution of $\mu$ with Pe computed numerically with GITT (blue circles).
			The red curve is the result of the fitting of the data with our toy-model Eq.~(\ref{eqmu2}).
			All curves correspond to $\varepsilon = 0.85$.}
			\label{figmufit}
		\end{figure}

		This model is tested against GITT data in Fig.~\ref{figmufit}.
		In the Newtonian regime, the two time scales collapse on each other, Pe $\propto$ Wi, and $\mu$ has the form of
		Eq.~(\ref{eqmu}).
		When going closer to the ideal granular glass transition, the two time scales separate, as in Fig.~\ref{figmuRecap},
		and $\mu^{(2)}$ gets closer to a constant, while $\mu^{(1)}$ still depends on Pe.
		Fitting the GITT data with such a model yields the red curves in Fig.~\ref{figmufit}.
		The agreement between the numerical data and the model is satisfactory.

	\subsection{The $\mu(I)$ regime}
	
		In the previous section, we did not discuss the last regime of dry granular flows: the Bagnold regime.
		In this regime it has been established experimentally that $\mu$ follows a phenomenological law, and depends on only
		one dimensionless quantity, the inertial number $\mathcal{I} = \dot{\gamma} d \sqrt{n/P}$ --- $n$ being the particle's
		density and $d$ their diameter --- which can be understood as the ratio of two time scales \cite{Cassar05}:
		the advection time scale $t_\gamma=1/\dot{\gamma}$, and the time scale of free-fall in a pressure field $P$, $t_{ff}$
		, which is the characteristic scale of the ballistic short time motion (an explicit derivation
		can be found in \cite{Cassar05}).
		The $\mu(\mathcal{I})$ law writes:
		\begin{equation}
		\label{eqmui}
			\mu(\mathcal{I}) = \mu_1 + \frac{\mu_2 - \mu_1}{1+\mathcal{I}_0/\mathcal{I}}\,,
		\end{equation}
		where $\mathcal{I}_0$, $\mu_1$ and $\mu_2$ are adjustable parameters.
		This law has been tested in a wide variety of flow geometries \cite{GDR04,Lagree11,Pahtz20}, and has proven to be
		successful, even in very recent experiments \cite{Forterre18,Tapia19}.
		This is crucial insofar as it means that the law Eq.~(\ref{eqmui}) provides knowledge about intrinsic properties of
		granular liquids.
		
		Let us examine the $\mu(\mathcal{I})$ law in the light of our toy-model.
		As explained before, in the Bagnold regime, the shear rate is strong enough so that the system lies in the advection
		dominated regime Wi$\gg1$.
		Therefore the long time scale ratio $\Gamma/\dot\gamma$ is very small, and the contribution $\mu^{(2)}$ is roughly constant.
		This is consistent with the fact that in typical experiments, the variation of $\mu$ over the whole $\mathcal{I}$
		range is mild --- it typically varies between $0.4$ and $0.65$.
		The subleading variations thus come from the change of short-time decay scale,
		that can be observed in Fig.~\ref{figSigmaRecap}.
		
		As we argued above, by definition, Pe is the ratio of the time scale associated with the short-time motion of the particle,
		which in this case can be identified with $t_{ff}$, with the advection time scale.
		Therefore, Pe$\propto \mathcal{I}$, as can be observed on the numerical data obtained by solving the full GITT equations displayed on Fig.~\ref{figIPe}.
		This argument is only qualitative, but a full demonstration of the link between the two quantities is tedious due to the presence
		of $P$ in $\mathcal{I}$, as no simple approximation of the equation of state is known to hold in all the studied regimes.
		
		Let us now examine a bit more in details the different regimes.
		The behavior of the system is controlled by three independent time scales: the advection time scale $t_\gamma$, the
		free fall time scale $t_{ff}$, and the scale of the structural relaxations $t_\Gamma$.
		In the Bagnold regime, the final relaxation is always controlled by advection, therefore $t_\gamma\ll t_\Gamma$.
		Hence, three different regimes can be defined depending on the values of $t_{ff}$:
		\begin{itemize}
			\item[(i)] $t_{ff}\ll t_\gamma \ll t_\Gamma$: Quasi-static regime. \\
				In this regime, $t_{ff}$ is the smallest time scale, $\mathcal{I}=t_{ff}/t_\gamma\ll 1$.
				From what we established before, Wi$\gg1$ and Pe$\ll1$.
				The effective friction, given by Eq.~(\ref{eqmu2}) is thus dominated by the long-time contribution
				$\mu^{(2)}$.
				By analogy with Eq.~(\ref{eqmui}), we can identify,
				\begin{equation}
				\label{eqQs}
					\mu \simeq \mu^{(2)} \simeq M_1^{(2)} = \mu_1\,.
				\end{equation}
				This corresponds to the black curve in Fig.~\ref{figmuRecap}, where the two relaxation time scales are
				clearly separated.
				
				Let us emphasize that our results only hold for moderately low values $\mathcal{I}\gtrsim 10^{-3}$,
				beyond that interparticle friction plays the dominant role, and defines the physics of the jamming
				transition \cite{Staron10,DeGiuli15,DeGiuli16,DeGiuli17b}.
				
			\item[(ii)] $t_\gamma\simeq t_{ff} \ll t_\Gamma$: $\mu(\mathcal{I})$ regime. \\
				This is the regime where $\mu$ varies between its two limiting values $\mu_1$ and $\mu_2$.
				In this regime, the two relaxation time scales get closer and closer to each other until they finally
				merge into one.
				The decay of $\Phi$ is controlled by $t_{ff}\simeq t_{\gamma}$.
				
			\item[(iii)] $t_\gamma\ll t_{ff}\simeq t_\Gamma$: Dilute liquid limit. \\
				In the limit of the lowest packing fractions accessible to the granular liquid phase, the advection
				time scale becomes even smaller than the internal relaxation time scale.
				In such a regime, both Pe$\gg1$ and Wi$\gg 1$.
				Comparing the toy-model $\mu$ Eq.~(\ref{eqmu2}) and the experimental law Eq.~(\ref{eqmui}) leads to:
				\begin{equation}
					\mu\simeq M_1^{(1)} + M_1^{(2)} = \mu_2 \,,
				\end{equation}
				which together with Eq.~(\ref{eqQs}) leads to $M_1^{(1)} = \mu_2-\mu_1$.
				
		\end{itemize}
		
		Finally, defining $\mathcal{I}_0= M_2^{(1)}\,\mathcal{I}/$Pe,
		\begin{equation}
			\mu^{(1)} = \frac{\mu_2-\mu_1}{1+\mathcal{I}_0/\mathcal{I}}\,,
		\end{equation}
		so that recalling that in the Bagnold regime Wi$\gg1$ is always true, this equation combined with Eq.~(\ref{eqQs})
		shows that Eq.~(\ref{eqmu2}) is exactly equivalent to Eq.~(\ref{eqmui}).

		It is also interesting to interpret the above results in terms of Maxwell's model.
		Since $\sigma=\int dt \dot{\gamma} G(t)$, we can identify $G(t)$ with $\Phi(t)^2$, which leads to the following
		equation for the time dependent shear modulus:
		\begin{equation}
		\label{eqG}
			G(t) = \left[(G_0-G_\infty)e^{-2\Gamma^{(1)}t} + G_\infty e^{-2\Gamma^{(2)}t}\right]
			e^{-(\dot{\gamma}t)^2/\gamma_c^2}\,,
		\end{equation}
		where, recalling that $G(t)$ follows a two-step decay similar to that of $\Phi(t)$, $G_0$ is the initial
		value of the shear modulus, and $G_\infty$ that of the plateau $1/\Gamma^{(1)}\ll t \ll1/\Gamma^{(2)}$.
		Accordingly, $G_0\geqslant G_\infty$.
		This model differs from the above non-linear Maxwell model because of two features: (i) it is expressed in terms of
		not only one but two characteristic shear moduli, which allows for a richer phenomenology, and (ii) the
		contribution of the advection time $\tau_\gamma=\gamma_c/\dot\gamma$ is now quadratic instead of linear.
		Consequently, the shear stress can be written:
		\begin{equation}
		\label{eqS3}
			\sigma = \frac{\sigma_y\left(\frac{G_0}{G_\infty}-1\right)}{1+\overline{\gamma}_c/\text{Pe}}
			+ \frac{\sigma_y}{1+\overline{\gamma}_c/\text{Wi}}\,,
		\end{equation}
		where $\sigma_y$ and $G_\infty$ are related by Eq.~(\ref{eqSy}).
		The pressure can also be decomposed:
		\begin{equation}
			P = P_0 + \frac{P_1}{1+\overline{\gamma}_c/\text{Pe}} + \frac{P_2-P_1}{1+\overline{\gamma}_c/\text{Wi}}\,,
		\end{equation}
		what finally leads to the following expressions for $\mu_1$ and $\mu_2$:
		\begin{equation}
			\mu_1 = \frac{G_\infty \overline{\gamma}_c}{P_0+P_1}\ ,\ \mu_2 = \frac{G_0 \overline{\gamma}_c}{P_0+P_2}\,.
		\end{equation}
		Hence, $\mu_1$ corresponds to the plateau elastic response of the viscoelastic fluid, whereas $\mu_2$ is associated
		with its initial value before the first step of the decay.
		Consistently with $G_0\geqslant G_\infty$, $\mu_1\leqslant \mu_2$ always holds.
		Note that the pressures appearing in these expressions are the limiting values of the pressure
		in the quasi-static (for $\mu_1$) and dilute liquid limit (for $\mu_2$).
		Since both limits correspond to regimes in which the role played by advection is crucial, the corrections to the
		hydrostatic pressure $P_0$ due to the shear are significant \cite{Coquand20}.

		Finally, the identification of $\mu_1$ and $\mu_2$ with the model of $\dot\gamma$-dependent friction of Savage and
		Hutter \cite{Savage89} (see details in appendix \ref{ASH}) allows us to identify $\mu_1=\tan(\delta_S)$ and
		$\mu_2=\tan(\delta_D)$, where $\delta_S$ and $\delta_D$ that delimit the regime in which steady shear flow can
		develop down slopes.
		
		All in all, in the light of the above discussion, our toy model can be rewritten in full generality as:
		\begin{equation}
		\label{eqmu3}
			\mu(\mathcal{I}, \text{Wi}) = \frac{\mu_1}{1 + M/\text{Wi}} + \frac{\mu_2 - \mu_1}{1 + 
			\mathcal{I}_0/\mathcal{I}}\,.
		\end{equation}
		Adding the constraint that in the Bagnold regime the final relaxation process is always controlled by shear advection, in which case
		Wi$\gg1$, this equation reduces exactly to the $\mu(\mathcal{I})$ law.
		It was discussed in a previous studies \cite{Peyneau08,Clavaud17,Pahtz20,Coquand20} that $\mu_1$ was not a direct consequence
		of the presence of interparticle friction, but also arose from collective effects.
		What the present study adds to this picture is the relation between the non-zero value of $\mu_1$ and the separation
		of time	scales in the relaxation of $\Phi$ towards 0.		

	\subsection{Granular suspensions}
	
		In a series of recent studies, striking similarities between the laws governing the flow of dry granular liquids and
		granular suspensions have been highlighted \cite{Courrech03,Cassar05,Boyer11,DeGiuli15,Guazzelli18,Tapia19,Pahtz19,Suzuki19}.
		For the sake of consistency, let us let aside the considerations about the regime close to the jamming transition 
		\cite{DeGiuli15,Tapia19}, and focus on the dense liquid regime.
		
		The main results can be summarized as follows:
		in presence of a viscous liquid, a new time scale related to the steady motion of particles submitted to a drag force
		proportional to its velocity, called $t_\eta$, must be taken into account \cite{Courrech03,Cassar05} (in the original paper \cite{Cassar05}
		another regime was considered where the drag force is proportional to the  square of the particle's velocity; This large Reynolds number regime
		is not considered here).
		The ratio of this time scale and the advection time scale defines a new dimensionless number
		$\mathcal{J}=\eta_\infty \dot{\gamma}/P$ (in the paper \cite{Cassar05} an additional coefficient related to the Darcy
		law was included in the definition of  $\mathcal{J}$; We chose to give here the most widely used notation),
		where in accordance with our previous notations $\eta_\infty$
		is the viscosity of the surrounding fluid.
		
		It was then observed that $\mu$ follows a law, similar to Eq.~(\ref{eqmui}), but where $\mathcal{J}$ rather
		than $\mathcal{I}$ plays the role of dimensionless number.
		More precisely, it was proposed in \cite{Boyer11} that $\mu(\mathcal{J})$ writes in the following form:
		\begin{equation}
		\label{eqmuS}
			\begin{split}
				\mu(\mathcal{J}) &= \mu^c(\mathcal{J})+\mu^h(\mathcal{J}) \\
					&= \mu_1 + \frac{\mu_2-\mu_1}{1+\mathcal{J}_0/\mathcal{J}}+\mu^h(\mathcal{J})\,.
			\end{split}
		\end{equation}
		In the above expression two kinds of terms are identified: a collisional contribution which form is very similar
		to Eq.~(\ref{eqmui}), and a hydrodynamic term that is tailored to reproduce Einstein's viscosity at low density.
		
		Let us examine this result in the light of our toy-model.
		As explained above, the main effect of the surrounding fluid is to introduce a new time scale $t_\eta$ that will
		compete with $t_{ff}$, $t_\Gamma$ and $t_\gamma$ to determine the leading behavior of $\mu$. 

		\begin{figure}
			\begin{center}
				\includegraphics[scale=0.55]{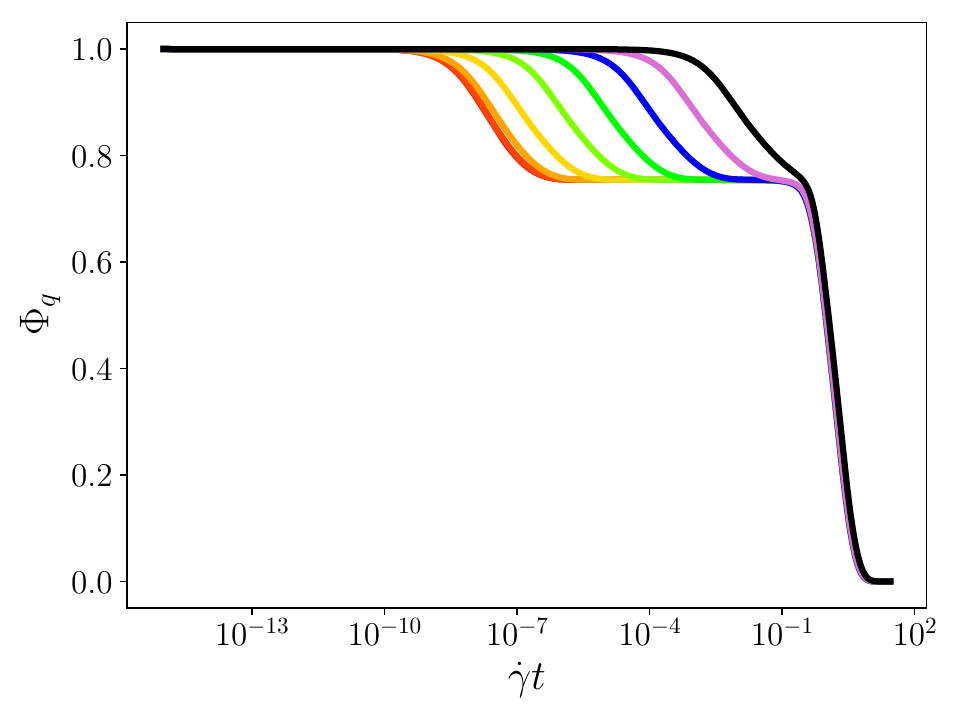}
			\end{center}
			\caption{Dynamical structure factor of granular suspensions for Pe=$10^{-9}$, $\varphi=0.58$ and Pe$_0$
			between $2.10^{-5}$ (red) and $2.10^{3}$ (black) obtained as a numerical solution of the GITT equations (\ref{eqRheo}).
			The curves for Pe$_0<10^{-5}$ all collapse on the red one.
			All curves correspond to $\varepsilon =0.85$.}
			\label{figGran}
		\end{figure}
		
		A good way to understand how the various time scales relate to each other is to first look at the numerical data from
		GITT.
		On Fig.~\ref{figGran} are displayed various profiles of $\Phi$.
		In order to visualize better the effect of $t_\eta$, which is a short-time scale, we choose a high $\varphi$ and a
		very low $\dot\gamma$, so that the long-time decay is delayed as much as possible.
		Two main regimes can be distinguished: if the Peclet number of the surrounding fluid Pe$_0=\dot{\gamma}d^2/D$ --- $D$
		being the diffusion coefficient in this fluid --- is small enough, all curves collapse as far as the first step of
		the decay is concerned, namely the first step of the decay is controlled by $\omega_c$ independent of the presence or
		absence of surrounding fluid.
		This is the dry granular liquid regime studied before, that extends to suspensions in a fluid with a low enough
		$\eta_\infty$.
		Then, for higher Pe$_0$'s, $t_\eta$ determines the scale associated with the first decay until it merges with $t_\Gamma$.
		This is the viscous suspension regime.

		\begin{table}
		\centering
			\begin{tabular}{|c|c|c|}
				\hline
				Short-time decay & Long-time decay & Flow regime \\
				\hline
				\hline
				\multirow{6}{2.0cm}{Dry~granular liquid: \vskip 0.2cm \hskip 0.2cm $t_\eta\ll t_{ff}$}
				& \multirow{3}{*}{$t_\gamma\gg t_\Gamma$} 
				& $t_\Gamma\simeq t_{ff}$: Newtonian \\[0.2cm]
				\cline{3-3}
				                                                          &
				& \multirow{2}{2.5cm}{$t_\Gamma\gg t_{ff}$ : Strongly coupled} \\[0.2cm] 
				 & &\\[0.2cm]
				\cline{2-3}
				                                                          & \multirow{3}{*}{$t_\Gamma \gg t_\gamma$}
				& $t_\gamma \gg t_{ff}$ : Quasi-static \\[0.2cm]
				\cline{3-3}
				                                                          &
				& $t_\gamma\simeq t_{ff}$ : Dense liquid \\[0.2cm]
				\cline{3-3}
				                                                          &
				& $t_\gamma \ll t_{ff}$ : Dilute liquid \\[0.2cm]
			 	\hline
				\hline
				\multirow{6}{2.2cm}{Granular suspension: \vskip 0.2cm \hskip 0.5cm $t_\eta\gg t_{ff}$} &
				\multirow{3}{*}{$t_\gamma\gg t_\Gamma$} 
				& $t_\Gamma\simeq t_\eta$: Newtonian \\[0.2cm]
				\cline{3-3}
				                                                          &
				& \multirow{2}{2.5cm}{$t_\Gamma\gg t_\eta$ : Strongly coupled} \\[0.2cm] 
				 & &\\[0.2cm]
				\cline{2-3}
				                                                          & \multirow{3}{*}{$t_\Gamma \gg t_\gamma$}
				& $t_\gamma \gg t_\eta$ : Quasi-static \\[0.2cm]
				\cline{3-3}
				                                                          &
				& $t_\gamma\simeq t_\eta$ : Dense liquid \\[0.2cm]
				\cline{3-3}
				                                                          &
				& $t_\gamma \ll t_\eta$ : Dilute viscous \\[0.2cm]
			 	\hline
			\end{tabular}
			\caption{Summary of the different regimes determined by the various time scales in granular suspensions.}
		\label{tabReg}
		\end{table}
		
		A summary of all the different regimes accessible to the system is given in Tab.~\ref{tabReg}.
		There are four competing time scales, but not all possible combinations are allowed.
		The short time decay is controlled either by $t_\eta$ or $t_{ff}$.
		When the liquid is Newtonian, which corresponds to memory effects playing a negligible role in the MCT equation
		Eq.~(\ref{eqMCT}), the short-time scale is equal to $t_\Gamma$.
		Note that by definition, $t_\Gamma\geqslant t_\eta,t_{ff}$.
		
		If $t_\Gamma\ll t_\gamma$, the long-time decay is independent of advection.
		Thus, the liquid can be either Newtonian, or strongly coupled if the density is high enough so that the cage effect
		becomes	important and the two relaxation scales separate from each other.
		
		The remaining regimes are the regimes controlled by advection, which can be either Bagnold or yielding.
		In the case $t_\eta \ll t_{ff}$, the $\mu(\mathcal{I})$ rheology is recovered.
		If $t_\eta\gg t_{ff}$ on the other hand, the short-time decay is determined by $\eta_\infty$.
		By a reasoning similar to the one we used previously, the ratio in Eq.~(\ref{eqmu2}) is
		thus a dimensionless number proportional to $1/\mathcal{J}$.
		This explains the strong similarities between the functional form of $\mu$ in the dry and suspended cases: what changes
		is simply the nature of the short-time scale; $\mu$ is still determined in the same fashion by the competition of a
		two-step decay profile, and advection.
		
		For example, let us consider the case of a dense granular suspension in the Bagnold regime.
		Strictly speaking, adding a viscous fluid changes the Bagnold equation since motion in the liquid adds a new source of
		energy dissipation in the system.
		However, the power dissipated by Stokes' force scales as $T$, whereas the power dissipated by collisions scales as
		$T^{3/2}$.
		Therefore, at high enough density, we can reasonably expect that collisions are the primary source of energy
		dissipation.
		Hence, the large time contribution $\mu^{(2)}$ should be unchanged compared to the dry case, that is: $\mu^{(2)}=\mu_1$.
		As for $\mu^{(1)}$, the only change is the nature of Pe which is now $\propto \mathcal{J}$.
		All in all,
		\begin{equation}
		\label{eqmuGran}
			\mu = \mu_1 + \frac{\mu_2-\mu_1}{1+\mathcal{J}_0/\mathcal{J}}\,,
		\end{equation}
		where $\mathcal{I}_0$ has been changed into $\mathcal{J}_0$ to account for the fact that the factor relating
		the original time scale ratio $t_\gamma/t_\eta$ to $1/\mathcal{I}$ or $1/\mathcal{J}$ may differ; but the other coefficients are unchanged.
		In particular, in Eq.~(\ref{eqmuGran}) only the collisional part $\mu^c$ contributes.
		This is consistent with the experimental findings of \cite{Cassar05}.
		It also means that the value of $\mu$ in the quasi-static limit should be the same in the dry and the suspended cases,
		which is also consistent with experiments \cite{Cassar05,Guazzelli18,Tapia19} (note however that some caution is required,
		indeed in the deep quasi-static regime, friction becomes important \cite{DeGiuli15} and could induce significant changes to the picture
		presented here).
		
		When going away from this particular case, $\mu^{(2)}$ acquires a non-trivial structure which should account for
		$\mu^h$ (the additional, higher order contributions to $\mu$ exhibited in appendix \ref{AITT} should also enter the 
		hydrodynamic component).
		In full generality, the three time scales toy-model predicts a rheological law of the form:
		\begin{equation}
		\label{eqReferee}
			\mu(\mathcal{J}, \text{Wi}) = \frac{\mu_1}{1 + M/\text{Wi}} + \frac{\mu_2 - \mu_1}{1 + 
			\mathcal{J}_0/\mathcal{J}}\,,
		\end{equation}
		where $M$ is the remaining constant.
		This equation shows a fundamental difference with Eq.~(\ref{eqmuS}): $\mu$ depends here on the two dimensionless numbers Wi and $\mathcal{J}$.
		This is a crucial lesson from our previous study of the dry case.
		If no liquid is present around the particles, however, in a typical experiment, the balance between the injected and the dissipated energies
		reduces to a balance between shear heating and dissipation by collisions, which then enforces the Bagnold scaling, so that only one of the
		two degrees of freedom remain.
		This is probably one of the reasons behind the large success of the $\mu(\mathcal{I})$ law.
		In granular suspensions, on the other hand, viscous drag is another important source of dissipation, so that both dimensionless numbers
		are independent, which explains why the rheology of suspensions has remained more elusive.

		We cannot however easily test our model Eq.~(\ref{eqReferee}) against the particular form of $\mu^{h}$  used in \cite{Boyer11} because (i) our
		toy-model expresses everything in terms of ratios of time scales, whereas \cite{Boyer11} fits a known
		$\varphi$-dependent function, and (ii) there is no guarantee that the low $\varphi$ limit of our model, built to be
		precise for $\varphi\gtrsim 0.4$,
		has the Einstein's viscosity as a natural limit as this expression is expected to be precise only up to
		$\varphi\simeq0.03$ \cite{Petford09}.
		
		Finally, an important feature of non-Brownian suspensions by opposition to dry granular liquids is the dilute liquid
		limit in which $\mu$ saturates in dry liquids, but continues to increase in suspensions \cite{Boyer11}.
		This goes a little bit beyond the frame of our model insofar as the regime in which $\mu$ really saturates is rarely
		reached by experiments, which means that the dense-liquid approach may break down before $\mu$ saturates, and the
		remaining variations can be accounted for by the difference between $\mathcal{I}_0$ and $\mathcal{J}_0$.
		Indeed, whereas in dry granular liquids when $\varphi$ is sufficiently decreased the stress is not well transmitted
		through the whole fluid, in the case of suspensions in a viscous enough liquid, the surrounding fluid can carry the
		stress to all particles and maintain the average velocity profile.
		Therefore, it is not even clear that the validity of our approach in the dilute limit extends to the same boundaries
		in the dry and suspended cases.

\section{Conclusion}

	To conclude, we adapted models used to describe the slow down of the dynamics of supercooled liquids and dense colloidal suspensions
	to the rheology of granular liquids, in the form of simple toy models that can be solved analytically, and yield
	constitutive equations that can be easily compared to numerical or experimental data.
	Because our model is not restrained to the Bagnold scaling regime of granular rheology, it shed the light on
	the importance of a third time scale, which was then crucial to make the connection between the rheology of dry granular
	flows and that of granular suspensions.
	The design of these toy models represent, in our opinion, a significant progress for the following reasons:
	(i) The constitutive equations we presented above are not purely phenomenological laws, but laws based on fundamental principles
	of liquid state physics; (ii) The toy model construction revealed that the complexity of the rheological behavior of granular liquids
	resides in the fact that it mixes physical phenomena occurring at different scales, both at the microscopic scale of individual particles' motion
	(via $t_{ff}$ and $t_\eta$), at the macroscopic scale of the material (via $t_\gamma$), but also at the mesoscopic scale of collective
	effects of a large number of particles (via $t_\Gamma$).
	In particular, the effect of $t_\Gamma$ was not anticipated in early experimental works, because most of the granular flows studied in the
	lab occur in the Bagnold regime where $t_\Gamma$ is much larger than any other time scale involved.
	However, $t_\Gamma$ remains crucial in the understanding of the bigger picture of granular rheology, where various flow regimes are allowed.
	(iii) The understanding of the fundamental processes at play in the different terms of the constitutive laws allows to encompass the rheology
	of both dry granular flows, and the flow of granular suspensions under the same formalism.
	This lead us to formulate non trivial predictions about the rheology of suspensions, such as Eq.~(\ref{eqReferee}), which
	can then be tested in experiments that can be conducted in simple shear cells like in \cite{Angelo23,Angelo25}, or in more
	refined rheometers like the one used in \cite{Boyer11}.


\section*{Acknowledgments}

	This work was funded by the Deutscher Akademischer Austauschdienst (DAAD) and the Deutsche Forschungsgemeinschaft (DFG),
	grant KR 48672.
	O. Coquand thanks K. Kelfoun for enlightening discussions.

\appendix

\section{Reduction of the ITT integrals}
\label{AITT}

	This appendix contains some details on the derivation of Eq.~(\ref{eqK0}) and Eq.~(\ref{eqK1}).
		
	First, we implement the approximation of the toy-model: we replace $\Phi_q(t)$ by an exponential decay, and add the
	advection Gaussian screening factor:
	\begin{equation}
		\begin{split}
			\mathcal{K}_0 &= \dot{\gamma}\int_0^{+\infty}dt\int_0^{+\infty}dq\frac{F_{1}(q,t)}{\sqrt{1+(\dot{\gamma}
			t)^2/3}} \\
			&\simeq \dot{\gamma}\int_0^{+\infty}dt \, e^{-2\Gamma t -2(\dot{\gamma}t)^2/\gamma_c^2} \\
			&= \frac{\gamma_c\sqrt{\pi}}{2\sqrt{2}}\,e^{\Gamma^2\gamma_c^2/(2\dot{\gamma}^2)}\text{erfc}
			\left(\frac{\Gamma\gamma_c}{\dot{\gamma}\sqrt{2}}\right)\,.
		\end{split}	
	\end{equation}
	Rigorously speaking, $\mathcal{K}_0$ includes an additional overall factor that accounts for the wave number integral.

\begin{figure}
		\begin{center}
			\includegraphics[scale=0.5]{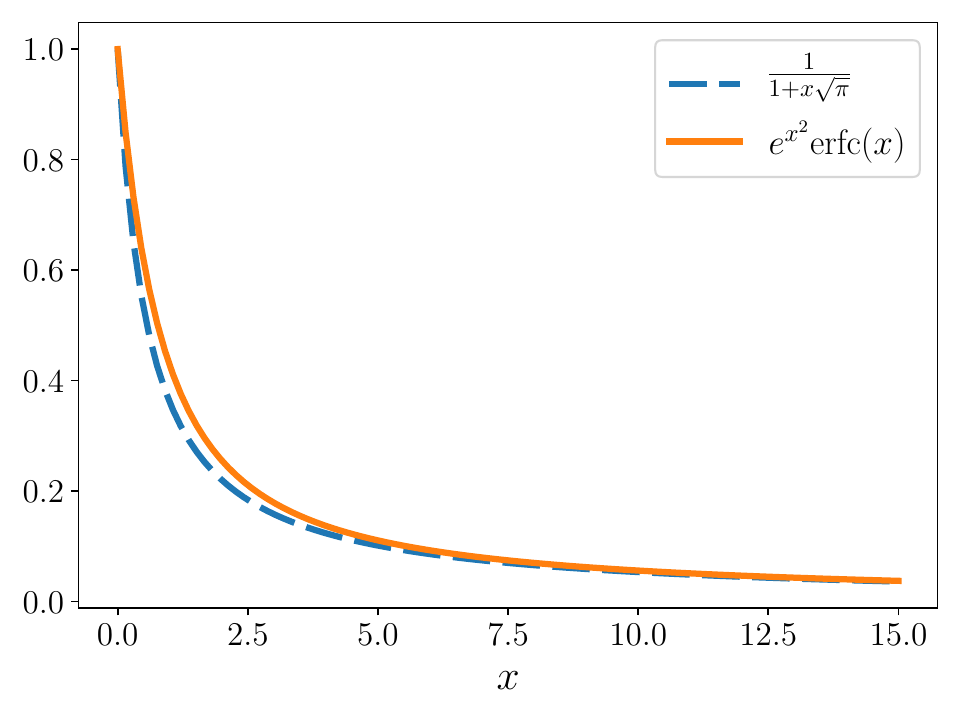}
		\end{center}
		\caption{Comparison of the functions $x\mapsto 1/(1+x\sqrt{\pi})$ (dashed line) and $x\mapsto e^{x^2}\text{erfc}(x)$
		(full line).}
		\label{figerfc}
	\end{figure}

	Then, we can replace $\exp(x^2)\text{erfc}(x)$ by $1/(1+x\sqrt{\pi})$ that shares the same $x\ll1$ and $x\gg1$ behaviors
	at leading order, and represents a satisfactory approximation of the whole function (see Fig.~\ref{figerfc}),
	which gives
	\begin{equation}
		\mathcal{K}_0 \simeq \frac{\overline{\gamma}_c}{2}\,\frac{1}{1+\overline{\gamma}_c\,u}\,.
	\end{equation}
	This procedure corresponds to representing $\mathcal{K}_0$ by its lowest order Pad\'e approximant (which is unique).

	The second ITT integrals appears in the expression of the correction to the pressure due to the shear (at our level of approximation
	it is not necessary to discriminate between $F_{1}$ and $F_{2}$):
	\begin{equation}
	\label{eqK11}
		\begin{split}
			\mathcal{K}_1 &= \dot{\gamma}\int_0^{+\infty}dt (\dot{\gamma} t)\int_0^{+\infty}dq
			\frac{F_{1}(q,t)}{\sqrt{1+(\dot{\gamma} t)^2/3}} \\
			& \simeq \frac{\gamma_c^2}{4}\left(1- \overline{\gamma}_c\,u\,e^{u^2\gamma_c^2/2}
			\text{erfc}\left(u\,\gamma_c/\sqrt{2}\right)\right)\,.
		\end{split}
	\end{equation}
	At this stage, the first option is to apply the same approximation as for $\mathcal{K}_0$, which gives the formula
	Eq.~(\ref{eqK1}).

	However, using this replacement for $\mathcal{K}_1$ is not as precise as it was for $\mathcal{K}_0$.
	Indeed, when $u\gg1$, the two leading order terms in the expression Eq.~(\ref{eqK11}) cancel against each other,
	so that $\mathcal{K}_1$ is $O(1/u^2)$, and not $O(1/u)$ as predicted by Eq.~(\ref{eqK1}).

	A better approximation can be built by relating $\mathcal{K}_1$ to $\mathcal{K}_0$:
	\begin{equation}
		\begin{split}
			\mathcal{K}_1 &\simeq \dot\gamma \int_0^{+\infty}dt(\dot{\gamma} t)\,e^{-2\Gamma t-2(\dot\gamma t)^2/
			\gamma_c^2} \\
			      &=-\frac{\dot\gamma}{2}\frac{d}{d\Gamma}\left(\int_0^{+\infty}\dot\gamma\,dt\,e^{-2\Gamma t-2(\dot\gamma t)^2/
			\gamma_c^2}\right) \\
			&=-\frac{\dot\gamma}{2}\frac{d\mathcal{K}_0}{d\Gamma}\\
			&=\frac{\overline{\gamma}_c^2}{4}\frac{1}{(1+\overline{\gamma}_c\,u)^2}\,.
		\end{split}
	\end{equation}
	This expression can be used to replace Eq.~(\ref{eqK1}) throughout the reasoning presented in this article.
	It yields higher order terms with higher powers of $\mathcal{I}$ in the $\mu(\mathcal{I})$ law, and therefore corresponds
	to a Pad\'e approximant of higher order.

	Given the excellent agreement between the $\mu(\mathcal{I})$ law and the available experimental data, we preferred to keep
	the simpler Eq.~(\ref{eqK1}) in our derivation.
	However, this computation reminds us that this law is only approximate.

\section{The non-linear Maxwell model}
\label{ANLMM}
	
		In their study of the rheology of colloidal suspensions \cite{Fuchs03}, Fuchs and Cates designed a simple toy-model
		which reproduces the ability of the complex colloidal liquid to interpolate between the Newtonian and the yielding
		regimes.
		In the Maxwell model of viscoelastic fluids in which the shear rate is decomposed into a solid-like and a liquid-like
		contributions, it can be established that the dynamical shear modulus $G(t)$ follows an evolution of type
		$G(t)=G_\infty e^{-t/\tau}$.
		In colloidal suspensions, as in the case of granular fluids, there are two characteristic time scales in competition:
		the structural relaxation time scale $\tau$, and the advection time scale $\tau_\gamma=c/\dot\gamma$
		($c$ being an unimportant constant).
		Fuchs and Cates then proposed to replace the time scale in the Maxwell model of viscoelasticity by $\tau_M$ defined as:
		\begin{equation}
			\frac{1}{\tau_M} = \frac{1}{\tau} + \frac{1}{\tau_\gamma}\,,
		\end{equation}
		what leads to the following expression for the shear stress:
		\begin{equation}
		\label{eqNLMM}
			\sigma= \dot{\gamma}\left(\eta_\infty + \frac{G_\infty\,\tau}{1+\dot{\gamma}\tau/c}\right)\,,
		\end{equation}
		where $\eta_\infty$ is the high shear limiting viscosity.
		
		The interpretation of Eq.~(\ref{eqNLMM}) goes as follows: at low density the structural relaxation
		time scale is small, so that, $\dot{\gamma}\tau\ll 1$ and $G_\infty\tau\ll\eta_\infty$, so
		that $\sigma\simeq\eta_\infty\dot{\gamma}$;
		whereas as the density increases, the internal relaxation becomes very slow, so that $\dot{\gamma}\tau\gg1$,
		and $\sigma\simeq G_\infty c$, which corresponds to a material of yield stress $\sigma_y^{FC}(c)=G_\infty c$.
		
		Since our toy-model also applies to colloidal suspensions after taking the elastic limit $\varepsilon\rightarrow1$,
		it is instructive to compare it to the non-linear Maxwell model of Fuchs and Cates.
		In our setup, the scale of structural relaxation is given by $\Gamma$, leading to the identification $\Gamma =  1/\tau$.
		From Eq.~(\ref{eqYield}), the yield stress corresponds to $\sigma_y$, so that $\sigma_y = G_\infty c$.
		From Eq.~(\ref{eqNew}), we can further identify $c$ and $\overline{\gamma}_c$.
		Plugging this back into Eq.~(\ref{eqSig}) yields:
		\begin{equation}
		\label{eqInt}
			\sigma = \frac{\sigma_y}{1 + \overline{\gamma}_c \Gamma/\dot\gamma}
			= \dot\gamma\,\frac{G_\infty\,\tau}{1+\dot{\gamma}\tau/c}\,,
		\end{equation}
		which is almost exactly identical to Eq.~(\ref{eqNLMM}), except for the first term.
		Note that such term does not derive naturally from the Maxwell model either, and has to be added afterwards.

		Indeed, as discussed above, the rheology is not governed by a competition between two, but three time scales.
		While for dense colloidal suspensions the main effects are described by the second term in Eq.~(\ref{eqNLMM}),
		as $\varphi$ decreases, the influence of the short-time decay of $\Phi$ become more and more important.
		In colloidal suspensions, the short-time dynamics is determined by the motion in the viscous fluid, with a
		time scale $\tau_\eta\propto\eta_\infty$.
		Furthermore, since the surrounding liquid is not supercooled, we can suppose that the short-time contribution
		$\sigma^{(1)}$ in the Newtonian regime, so that, according to Eq.~(\ref{eqNew}), $\sigma^{(1)}=\eta_\infty \dot\gamma$.
		Finally, with $\sigma^{(2)}$ given by Eq.~(\ref{eqInt}), the initial model of Fuchs and Cates Eq.~(\ref{eqNLMM})
		is recovered.		

\section{The Savage and Hutter model}
\label{ASH}

	In \cite{Savage89}, Savage and Hutter proposed a model of $\dot\gamma$-dependent friction, defined in terms of two
	universal functions $f_1(\varphi)$ and $f_2(\varphi)$ that writes:
	\begin{equation}
	\label{eqSH}
		\mu^{SH} = \tan(\delta) = \frac{P_0(\varphi)\tan(\delta_S) + f_2(\varphi)\dot\gamma^2}{P_0(\varphi)+
		f_1(\varphi)\dot\gamma^2}\,,
	\end{equation}
	where $\delta_S$ is the minimal angle for a steady flow to be sustained on a given slope.
	The expression Eq.~(\ref{eqSH}) is justified as follows: the numerator is the shear stress that can be decomposed as a
	yield stress that survives in the limit $\dot\gamma\rightarrow0$, $\sigma_y^{SH} = P_0 \tan(\delta_S)$, and a correction
	that typically goes as $\dot\gamma^2$ in the Bagnold regime. The denominator is nothing but the similar expression
	for the pressure.

	In order to make the comparison with our model easier, let us forget about the $\varphi$ dependence, and introduce the
	following coefficients: $\mu_1^{SH}=\tan(\delta_S)$ --- the effective friction coefficient in the $\dot\gamma\ll1$
	regime --- $\alpha = f_1/P_0$ and $\mu_2^{SH} = f_2/f_1$.
	Eq.~(\ref{eqSH}) can thus be rewritten:
	\begin{equation}
	\label{eqSH2}
		\begin{split}
			\mu^{SH} &= \frac{\mu_1^{SH}}{1 + \alpha\,\dot\gamma^2} + \frac{\mu_2^{SH}}{1+1/(\alpha\dot\gamma^2)}\\
			&= \mu_1^{SH} + \frac{(\mu_2^{SH}-\mu_1^{SH})(1+\alpha\dot\gamma^2)}
			{2+\alpha\dot\gamma^2 + 1/(\alpha\dot\gamma^2)}\,.
		\end{split}
	\end{equation}
	This expression describes an evolution qualitatively similar to $\mu(\mathcal{I})$ between two finite limits $\mu_1$
	and $\mu_2$ when $\dot\gamma$ is varied.
	Since Eq.~(\ref{eqSH2}) is written in the Bagnold regime, the possibility to have a Newtonian fluid as in our toy-model
	is excluded.

	Finally, in the model of Savage and Hutter, $\mu_1^{SH}$ and $\mu_2^{SH}$ define two friction angles $\delta_S$ and
	$\delta_D$ that separate different flow regimes down a slope of angle $\zeta$:
	(i) if $\zeta<\delta_S$, the flow stops at some point because of the friction inside the complex fluid; (ii) if
	$\delta_S\leqslant\zeta\leqslant\delta_D$, the fluid reaches a steady flow regime if let to flow for a long enough time;
	(iii) if $\zeta>\delta_D$, the flow is continuously accelerated.
	Note that $\delta_S\leqslant\delta_D$ is consistent with $\mu_1\leqslant\mu_2$.

\section{Mode Coupling Equations for inelastic hard spheres}
\label{AMCT}

	This appendix merely summarizes the equations.
	Their derivation can be found in the paper\cite{Kranz20}.

	The dynamics of $\Phi_q(t)$ is given by a Mori-Zwanzig type equation~:
	\begin{equation}
		\begin{split}
			\ddot{\Phi}_q(t) &+ \nu_{q(t)}\dot{\Phi}_q(t)+q^2(t) C_{q(t)}^2 \Phi_q(t) \\
			+& q^2(t) C_{q(t)}^2 \int_0^td\tau\,m_q(t,\tau)\dot{\Phi}_q(\tau) =  0 \:.
		\end{split}
	\end{equation}

	In this equation, the characteristic frequencies are~:
	\begin{equation}
		\nu_q = \frac{1+\varepsilon}{3} \,\omega_c\,\Big[1+3j_0''(qd)\Big]\:,
	\end{equation}
	(where $d$ is the particle's diameter and $j_0$ is the zeroth-order spherical Bessel function),
	and $\Omega_q^2=q(t)^2C_{q(t)}^2$, with the speed of sound $C_q$ expressed as~:
	\begin{equation}
		C_q^2 = \frac{T}{S_q}\left[\frac{1+\varepsilon}{2}+\frac{1-\varepsilon}{2}\,S_q\right]\:.
	\end{equation}
	The mode-coupling kernel $m_{q}$ is quite similar to its well-known value in the elastic limit $\varepsilon\rightarrow1$,
	\begin{equation}
	\label{eqmMCT}
		\begin{split}
			m_q(t,\tau) &
			= A_{q(t)}(\varepsilon)\frac{S_{q(t)}}{nq^2}\int\frac{d^3k}{(2\pi)^3} S_{k(\tau)}S_{p(\tau)} \\
			&\times\big[(\hat{\mathbf{q}}.\mathbf{k})nc_{k(t)} + (\hat{\mathbf{q}}.\mathbf{p})nc_{p(t)}\big]\\
			&\times\big[(\hat{\mathbf{q}}.\mathbf{k})nc_{k(\tau)} + (\hat{\mathbf{q}}.\mathbf{p})nc_{p(\tau)}\big] \\
			&\times\Phi_{k(\tau)}(t-\tau)\Phi_{p(\tau)}(t-\tau) \,.
		\end{split}
	\end{equation}
	In this equation, $n$ is the fluid's density, hats denote normalized vectors, $c_q$ denote the direct correlation function,
	and $A_q(\varepsilon)$ is a prefactor given by \cite{Kranz13}~:
	\begin{equation}
	\label{eqA}
		A_q^{-1}(\varepsilon) = 1+\frac{1-\varepsilon}{1+\varepsilon}\,S_q\:,
	\end{equation}
	which does equal to 1 in the elastic limit, as required by consistency.

\bibliography{TM.bib}
\end{document}